\documentclass[useAMS,usenatbib]{mn2e}

\usepackage{amsmath} 
\usepackage{amssymb} 
\usepackage{graphicx}

\newcommand\ion[2]{#1$\;${\scshape{#2}}}
\newcommand{\fixbib}[1]{}

\title[$z \sim 6$ metal-line absorbers]{Testing the effect of galactic feedback on the IGM at $z \sim 6$ with metal-line absorbers}

\author[L.C. Keating et al.]{\parbox{\textwidth}{Laura C. Keating$^{1,2}$\thanks{E-mail:
    lck35@ast.cam.ac.uk}, Ewald Puchwein$^{1,2}$, Martin G. Haehnelt$^{1,2}$, Simeon Bird$^{3}$ and James S. Bolton$^{4}$}\vspace{0.4cm} \\
\parbox{\textwidth}{$^1$Institute of Astronomy, University of Cambridge,
  Madingley Road, Cambridge, CB3 0HA, UK\\
$^2$Kavli Institute  for Cosmology,  University of Cambridge,
  Madingley Road, Cambridge, CB3 0HA, UK\\
$^3$Department of Physics and Astronomy, Johns Hopkins University, Baltimore, MD 21218, USA\\
$^4$School of Physics and Astronomy, University of Nottingham, University Park, Nottingham, NG7 2RD, UK}
}

\begin{document}

\date{Accepted 2016 May 27. Received 2016 May 27; in original form 2016 March 10\\
This is a pre-copyedited, author-produced version of an article accepted for publication in MNRAS following peer review.}

\pagerange{\pageref{firstpage}--\pageref{lastpage}} 

\pubyear{2015}

\maketitle

\label{firstpage}

\begin{abstract}
We present models of low- and high-ionization metal-line absorbers (\ion{O}{i}, \ion{C}{ii}, \ion{C}{iv} and \ion{Mg}{ii})  during the end of the reionization epoch, at $z \sim 6$. Using four cosmological hydrodynamical simulations with different feedback schemes (including the Illustris and Sherwood simulations) and two different choices of hydro-solver, we investigate how the overall incidence rate and equivalent width distribution of metal-line absorbers varies with the galactic wind prescription. We find that the \ion{O}{i} and \ion{C}{ii} absorbers are reasonably insensitive to the feedback scheme. All models, however, struggle to reproduce the observations of \ion{C}{iv} and \ion{Mg}{ii}, which are probing down to lower overdensities than \ion{O}{i} and \ion{C}{ii} at $z \sim 6$, suggesting that the metals in the simulations are not being transported out into the IGM efficiently enough. The situation is improved but not resolved if we choose a harder (but still reasonable) and/or (locally) increased UV background at $z \sim 6$. 
\end{abstract}


\section{Introduction}

Observations of metal absorption lines in the spectra of QSOs out to $z \gtrsim 6$ \citep[e.g.,][]{ryanweber2009,becker2011,simcoe2011,dodorico2013} are providing an important probe into the enrichment and ionization state of the high-redshift intergalactic medium (IGM). Ions with atomic transitions redward of Ly$\alpha$ (Ly$\alpha$) provide the opportunity to study the metal-enriched IGM in absorption, even after the saturation of the Ly$\alpha$ forest at $z \sim 6$ \citep[see][for a recent review on quasar absorption lines during the reionization epoch]{becker2015rev}. Detections of different ions can provide complementary information, as high- and low-ionization lines tend to trace gas at different densities and temperatures. In this work, we focus our efforts on modelling four ions: \ion{O}{i}, \ion{C}{ii},  \ion{C}{iv} and \ion{Mg}{ii} as \citet{becker2011}, \citet{matejek2012} and \citet{dodorico2013} have published equivalent widths for absorption systems containing these ions out to $z \sim 6$.  Understanding detections of \ion{O}{i} is particularly useful for observations pushing into the reionization epoch, as \ion{O}{i} can provide an important analogue for the presence of neutral hydrogen due to their similar ionization energies \citep{oh2002}.

Modelling these absorption systems is a non-trivial task, however, due to the uncertainties in the shape and amplitude of the UV background at $z \sim 6$ as well as to which density the IGM is enriched with metals \citep{oppenheimer2009}. High-resolution simulations are also required to accurately resolve the small self-shielded systems responsible for hosting the neutral gas \citep{bolton2013}. Previous works have also explored the effect of varying models of galactic outflows \citep{oppenheimer2006}, changing the initial mass function for Population III stars \citep{pallottini2014a} and self-consistently simulating the hydrodynamics and ionization state of the gas \citep{finlator2013,finlator2015}.

At intermediate redshifts ($z = 2-4$), there have been a number of efforts to test the sensitivity of metal-line absorbers (mostly \ion{C}{iv}) to the feedback prescription. \citet{oppenheimer2006, oppenheimer2008} and \citet{oppenheimer2012} have had success matching observations of the mass density of \ion{C}{iv} across cosmic time using a model where the velocity and mass-loading of the winds scales with properties of the host galaxy. \citet{tescari2011} compared a range of feedback models to \ion{C}{iv} absorbers from \citet{dodorico2010} and showed that, while properties of \ion{H}{i} are robust to the choice of feedback prescription, models of \ion{C}{iv} are much more sensitive to the feedback scheme. \citet{suresh2015} and  \citet{bird2015b} explored the effect of varying the feedback prescription on the circumgalactic medium (CGM) and properties of \ion{C}{iv} absorption systems, confirming that metal-line absorbers are very constraining on feedback implementations. \citet{rahmati2015} have also explored the evolution of high-ionization metal absorbers in the EAGLE simulation \citep{schaye2015} from $z=0-6$, finding generally good agreement but with a suggestion that their models are failing to reproduce the abundances of the high-redshift absorbers.

The choice of feedback model is clearly critical in determining the enrichment level and temperature of the $z \sim 6$ IGM and CGM. Many feedback models are chosen so as to reproduce properties of $z=0$ galaxies, such as the galaxy stellar mass function, and it is therefore interesting to test them against statistics they were not tuned to match over a range of redshifts. Metal-line absorbers provide an interesting constraint on the efficiency of galactic winds in transporting metals and allow us to test these models using some of the highest redshift observations available. These high-redshift tests are extremely important, as models that fail to reproduce observations at $z \sim 6$ will also be incorrect at lower redshifts and may match lower redshift observations for the wrong reasons. In this paper we use a range of simulations, using both different hydrodynamical codes and different feedback schemes to explore the enrichment of the $z \sim 6$ IGM and investigate how this changes with different models of galactic winds. These include the publicly available Illustris simulation \citep{genel2014,vogelsberger2014,vogelsberger2014b,nelson2015} run with the moving mesh code \textsc{arepo} \citet{springel2010arepo}, as well as some simulations containing variants on Illustris physics previously described in \citet{bird2014,bird2015,bird2015b}. We also use a run containing supernova feedback from the Sherwood Simulation Suite (Sherwood), designed for high-resolution studies of the Ly$\alpha$ forest in large volumes \citep{bolton2016}. This project aims to bridge the important gap between small and large scales with a suite of some of the highest resolution Ly$\alpha$ forest simulations performed to date within large volumes.

The structure of this paper is as follows. In the second section, we discuss our different simulations and the effect of changing the feedback on the distribution of the metals. In the third section, we describe our ionization models and synthetic spectra we generate. In the fourth section, we make a comparison between our models and the observations and present some investigations into what may need to be changed in future work. We also make some predictions for what may be observed at $z > 6$. Finally, in the last section we present our conclusions.

\section{Models of Metal Absorption Lines}

\subsection{Hydrodynamical simulations}

\begin{table*}
\centering
\begin{tabular}{c|c|c|c|c|c|c}
Name & Code & Box Size (cMpc) & $m_{\textnormal{\scriptsize{dm}}} (M_{\odot})$ & $m_{\textnormal{\scriptsize{gas}}} (M_{\odot})$ & UV Background & $(H_{0}, \Omega_{\textnormal{\scriptsize{m}}}, \Omega_{\Lambda})$\\
\hline
Illustris & \textsc{arepo} & 106.5 & $6.3 \times 10^6$ & $1.3 \times 10^6$ & \citet{fauchergiguere2009} & (0.704, 0.273, 0.727)\\
Sherwood & \textsc{p-gadget3} & 59.0 & $7.9 \times 10^5$ & $1.5 \times 10^5$ & \citet{haardtmadau2012} & (0.678, 0.308, 0.692) \\
FAST \& HVEL & \textsc{arepo} & 35.5 & $1.0 \times 10^7$ & $2.1 \times 10^6$ & \citet{fauchergiguere2009} & (0.704, 0.273, 0.727) \\
\end{tabular}
\caption{Parameters of the simulations we use in this work: code that was used to run the simulation, the mass of the dark matter particle ($m_{\textnormal{\scriptsize{dm}}}$), the mass of the gas particle/resolution element ($m_{\textnormal{\scriptsize{gas}}}$), the UV background used and the cosmology used. In the case of the \textsc{arepo} simulations, we quote the mean gas mass of all resolution elements in the snapshot when we reference $m_{\textnormal{\scriptsize{gas}}}$.}
\label{simulations}
\end{table*}

\begin{table*}
\centering
\begin{tabular}{c|c|c|c|c|c}
Name & Wind model &  AGN & Self-shielding &  Notes \\
\hline
Illustris & \citet{vogelsberger2013} &  Yes  & Yes&  -\\
Sherwood & \citet{Puchwein2013gsmf} &  No  & No & -\\
HVEL & \citet{vogelsberger2013} &  Yes   & Yes& $v_{\textnormal{\scriptsize{wind,min}}}$ = 600 km s$^{-1}$ \\
FAST & \citet{vogelsberger2013} &  Yes  & Yes& 1.5 $\times$ Illustris wind vel\\
\end{tabular}
\caption{Further parameters of simulations: reference to the wind model used, whether AGN are included, whether the UV background is attenuated due to self-shielding during the hydrodynamical simulation and additional notes on the feedback schemes.}
\label{feedback}
\end{table*}

As we wish to explore the effect that feedback models have on the abundances of metal line absorbers, we use outputs from several different simulations, described below and summarised in Tables \ref{simulations} and \ref{feedback}. We use here output from the highest-resolution run of the Illustris simulation suite, Illustris-1, which we refer to as Illustris throughout. We also use simulations with variants of the Illustris feedback model, previously described in \citet{bird2014,bird2015,bird2015b}, where they were used to test the effect of feedback on DLAs and to look at \ion{C}{iv} absorbers at $z=2-4$. We also analyse a run using \textsc{gadget-3} from the Sherwood Simulation Suite that contains supernova feedback only. Unless otherwise stated, we use snapshots at $z=5.85$ for the \textsc{arepo} runs and at $z=6$ for the  \textsc{gadget-3}. This is to ensure that reionization is complete for both codes, as they were run with different UV backgrounds, and therefore that the gas temperatures are similar in both cases. We investigate the effect of changing the resolution of these simulations in Appendix \ref{sec:restest}.

\subsubsection{Codes}

The Illustris simulation was run using the moving mesh code \textsc{arepo} \citep{springel2010arepo}. The Sherwood Simulation Suite \citep{bolton2016} was run using the SPH code \textsc{gadget-3}, last described in \citet{springel2005gadget}. Both codes use a gravity solver based on the \textsc{tree-pm} scheme. For the purposes of this work, perhaps the most important difference between the two hydrodynamic solvers will be whether or not there is a possibility for gas of different metallicity to mix. In SPH codes, metals are assigned to a certain particle and cannot be transferred to other resolution elements. In mesh codes like \textsc{arepo}, however, metals can be advected from cell to cell. \textsc{arepo} also better resolves fluid instabilities that promote mixing. The result is that simulations run with SPH can predict higher metallicities in dense regions. As the \ion{C}{iv} fraction especially is expected to be largest in overdensities close to the mean at $z \sim 6$, this could be important for our results. 

\subsubsection{Galactic Winds and AGN Feedback}

All of these models make use of variable wind speed prescriptions \citep[e.g.,][]{oppenheimer2006,Puchwein2013gsmf,vogelsberger2014} that scale with some property of the galaxy or dark matter halo. The wind speed, $v_{\textnormal{\scriptsize{wind}}}$, here refers to the velocity with which particles are ejected from the interstellar medium (ISM). A variable-winds approach has been shown to reproduce the shape of the galaxy stellar mass function at $z=0$. Since the mass-loading factor $\eta_{\textnormal{\scriptsize{wind}}}$ scales like $v_{\textnormal{\scriptsize{wind}}}^{-1}$ or  $v_{\textnormal{\scriptsize{wind}}}^{-2}$ (depending on if the outflow is momentum- or energy-driven), this model suppresses star-formation most efficiently in low-mass galaxies.

The physics of the Illustris feedback model is described and tested in detail in \citet{vogelsberger2013} and \citet{torrey2014}. In this model, the galactic winds scale with the one-dimensional dark matter velocity dispersion: $v_{\textnormal{\scriptsize{wind}}}  = \kappa_{\textnormal{\scriptsize{wind}}} \sigma_{\textnormal{\scriptsize{DM}}}^{\textnormal{\scriptsize{1D}}}$.$\kappa_{\textnormal{\scriptsize{wind}}}$ is a dimensionless model parameter, set to 3.7 for this simulation. The dimensionless mass-loading factor is given by
\begin{equation}
\eta_{\textnormal{\scriptsize{wind}}} = 2 \times \left(\frac{974\, \textnormal{km s}^{-1}}{v_{\textnormal{\scriptsize{wind}}}}\right)^{2}.
\end{equation}
We also look at two simulations with variations of the Illustris wind physics. The HVEL model uses the same wind prescription as Illustris, but imposes a minimum wind speed of 600 km s$^{-1}$. This minimum wind speed corresponds to a halo mass of about $3 \times 10^{11} M_{\odot}$ at $z=6$ in the Illustris variable wind model. However, at $z \sim 6$ there are not yet any haloes with velocity dispersions or halo masses above this point. The FAST model has winds with velocities a factor of 1.5 times the wind speeds in the Illustris feedback model. This is achieved by changing the scaling factor between the wind velocity and dark matter velocity dispersion, i.e., setting $\kappa_{\textnormal{\scriptsize{wind}}}$ to 5.5. We do not consider here the WARM winds model, which \citet{bird2015b} found to match well the properties of $z=2-4$ \ion{C}{iv} absorbers, as it gave similar results to the FAST model at higher redshifts.

The Sherwood simulation uses the variable wind speeds model from \citet{Puchwein2013gsmf}. In this model, the wind velocities scale with the escape velocity of the galaxy: $v_{\textnormal{\scriptsize{wind}}} = 0.6 v_{\textnormal{\scriptsize{esc}}}$. The escape velocity of a gas particle is calculated by measuring the mass of the friends-of-friends group of which it is a part. Then the escape velocity is estimated by assuming the shape of the halo can be described using a Navarro-Frenk-White profile \citep{navarro1996}, such that
\begin{equation}
\label{eqn}
v_{\textnormal{\scriptsize{esc}}} = v_{\textnormal{\scriptsize{200}}} \times \sqrt{\frac{2c}{\ln(1+c) - \frac{c}{1+c}}}
\end{equation}
where $v_{\textnormal{\scriptsize{200}}}$ is the circular velocity at the virial radius, $r_{\textnormal{\scriptsize{200}}}$, and $c$ is the halo's concentration. This is estimated using the mass-concentration relation from \citet{neto2007}. The mass-loading is also determined by assuming the winds are energy-driven, and is set to
\begin{equation}
\label{eqn:p13}
\eta_{\textnormal{\scriptsize{wind}}} = 2 \times \left(\frac{958 \, \textnormal{km s}^{-1}}{v_{\textnormal{\scriptsize{wind}}}}\right)^{2}.
\end{equation}

All of these models use decoupled wind schemes, where wind particles are decoupled from hydrodynamical forces after their launch until they have reached some density threshold or a certain time has passed \citep[e.g.][]{springel2003}. The particles still interact gravitationally and radiative cooling is left on. This density threshold is set to 10 per cent of the star-formation threshold, which is $\rho_{\textnormal{\scriptsize{SF}}} = 0.13$ cm$^{-3}$ for the \textsc{arepo} simulations and $\rho_{\textnormal{\scriptsize{SF}}} = 0.33$ cm$^{-3}$ for the Sherwood simulation. By looking at volume-weighted spherically-averaged density profiles of haloes, we estimate that these density thresholds correpond to a average distance of $0.7 R_{\textnormal{\scriptsize{vir}}}$ for the  \textsc{arepo} simulations and $0.4 R_{\textnormal{\scriptsize{vir}}}$ for the Sherwood simulation. There is of course some scatter between individual haloes, however.

The Illustris, HVEL and FAST simulations all include AGN feedback, but the Sherwood simulation does not include black holes. We do not however expect this to change our results significantly (see, for example, \citet{tescari2011} and \citet{suresh2015}, who find that AGN are not very important for metal enrichment at high redshift).

\subsubsection{UV Background}

Illustris, HVEL and FAST were all run using the \citet{fauchergiguere2009} model of the UV background. A correction for the self-shielding of gas from the UV background is also incorporated into these simulations, based on the \citet{rahmati2013} model, however this is only included below $z=6$. The Sherwood simulation was run using the \citet{haardtmadau2012} UV background, which turns on at $z = 15$. Note that throughout this work, unless specified, we used the \citet{haardtmadau2012} model of the UV background for computing the ionization state of the gas. The original UV background of the simulations will still be important when considering, for example, the temperature of the gas in the simulations. We do investigate the effect of changing the shape and amplitude UV background on the metal-line absorbers we model in Section \ref{sec:shapeUV}. Ideally we would use a more realistic inhomogeneous UV background, such as in \citet{finlator2015}, but it is reassuring in that work they find that there is only a factor of $\sim 2$ difference between the neutral fractions calculated using their model and the \citet{haardtmadau2012} background. 

\subsection{Feedback and the High-z IGM/CGM}

\begin{table}
\centering
\begin{tabular}{c|c|c}
Name & $\rho_{\textnormal{\scriptsize{met}}} (10^{10} M_{\odot}/$cMpc$^{3})$& $\rho_{*} (10^{10} M_{\odot}/$cMpc$^{3})$ \\
\hline
Illustris & $9.8 \times 10^{-6}$ & $3.6 \times 10^{-4}$\\
Sherwood & $1.5 \times 10^{-5}$ & $7.9 \times 10^{-4}$\\
HVEL & $1.9 \times 10^{-5}$ & $5.9 \times 10^{-4}$\\
FAST & $9.2 \times 10^{-6}$ & $3.1 \times 10^{-4}$\\
\end{tabular}
\caption{Total mass density of metal-enriched gas ($\rho_{\textnormal{\scriptsize{met}}}$) and total stellar mass density ($\rho_{*}$) in the four simulations.}
\label{total_met_star}
\end{table}

\begin{figure*}
\includegraphics[width=2.1\columnwidth]{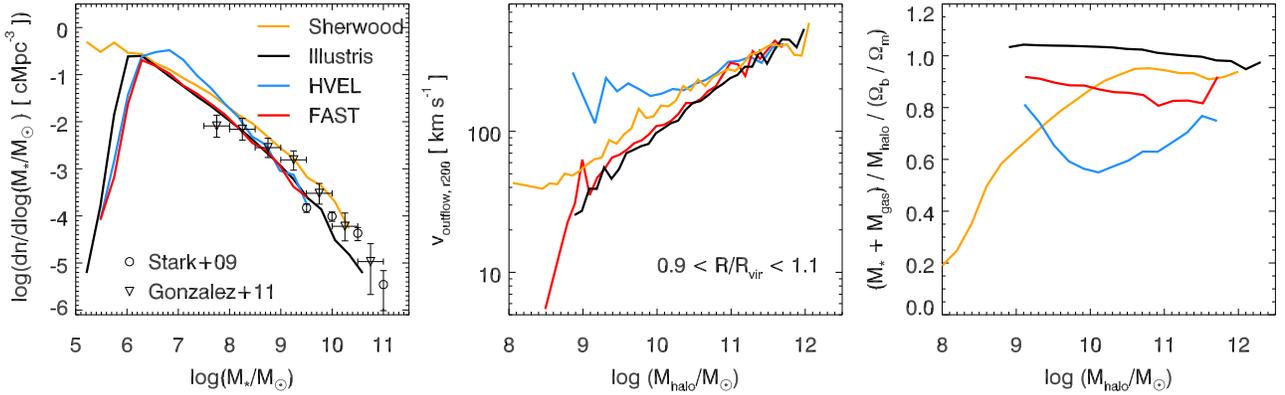}
\caption{Left: The stellar mass function at $z \sim 6$, compared with measurements from \citet{stark2009} and \citet{gonzalez2011}. Middle: Metallicity-weighted mean outflow velocity of particles in a shell at the virial radius as a function of halo mass. Right: The baryon fraction, normalised by the cosmic baryon fraction, as a function of halo mass, which we take to be the total mass of the FoF group.}
\label{vel_gasfrac_gsmf}
\end{figure*}

Galactic feedback is generally invoked to reconcile the shapes of the dark matter halo mass function and galactic stellar mass function \citep[e.g.,][]{white1991}. For our purposes, the feedback is important both as a means of regulating the number of stars produced (and hence the amount of metals) and also for distributing these metals throughout the IGM. In the left panel of Figure \ref{vel_gasfrac_gsmf}, we show the galaxy stellar mass function of the different simulations at $z=6$ for haloes resolved by at least 100 dark matter particles. Here we compare to observations of the stellar mass functions at $z \sim 6$ \citep{stark2009,gonzalez2011} but note that the Illustris run has already been compared with observations across a wide range of redshifts $z=0-7$ in \citet{genel2014}. The HVEL model shows an increase in galaxies with low stellar masses, as the constant fast winds provide a mass-loading factor that is too low to effectively flatten the lower end of the stellar mass function. It is also interesting to note that the \citet{Puchwein2013gsmf} model produces more stars than the \citet{vogelsberger2013} model at $z=6$, the opposite to what is seen at $z=0$ \citep{schaye2015}. We also find that the mass density of metals produced varies by about a factor of two among the four simulations, increasing in proportion with the total stellar mass (the values are given in Table \ref{total_met_star}).

In the middle panel of Figure \ref{vel_gasfrac_gsmf}, we show the metallicity-weighted mean outflow velocity (positive radial velocity), $v_{\textnormal{\scriptsize{outflow, r200}}}$, measured in a shell placed at the virial radius (at $0.9 < r/r_{\textnormal{\scriptsize{vir}}} < 1.1$). This outflow velocity is not the velocity kick given to wind particles, although the two are related. This was measured for a sample of haloes and the median in different halo mass bins is presented here. There is significant scatter among haloes though. We find that the trend in outflow velocity against halo mass is flattest for the HVEL model, as one would expect as it has a constant wind speed for low-mass haloes. Despite having similar wind prescriptions, the Sherwood simulation appears to have a higher outflow velocity than the Illustris simulation over a range of halo masses.

The baryon fraction inside the haloes varies quite considerably (right panel of Figure \ref{vel_gasfrac_gsmf}), with the models that contain faster winds generally having lower baryon fractions. There also seems to be a difference between the shape of the baryon fraction-halo mass relation between the simulations run with \textsc{arepo} and \textsc{gadget}.  The baryon fraction increases as a function of halo mass in the Sherwood simulation, and is rather flat across all halo masses in Illustris and the FAST simulations. This is in part due to the choice of UV background. The Sherwood simulation was run with a UV background that turns on at $z=15$ and results in hotter gas and therefore photoevaporation of some haloes with $\log (M_{\textnormal{\scriptsize{halo}}}/M_{\odot}) < 10$. We note also that these lines are only the mean baryon fractions and that there is scatter among the baryon fractions of individual haloes for a given mass.

\begin{figure*}
\includegraphics[width=2.1\columnwidth]{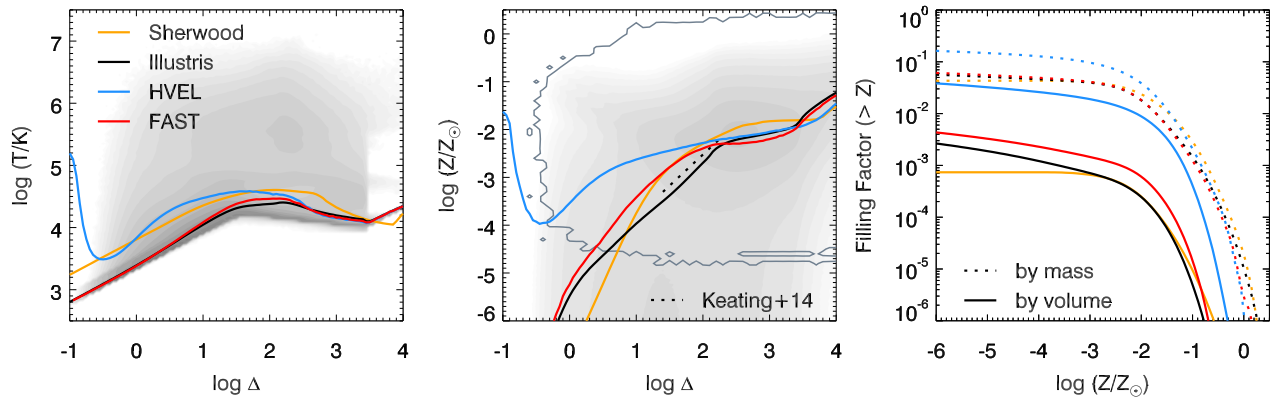}
\caption{Left: mean mass-weighted temperature in the four simulations as a function of gas overdensity. The grey shaded region shows the full temperature-density phase diagram for Illustris. Middle: mean mass-weighted metallicity in the four simulations as a function of gas overdensity. The grey shaded region shows the full metallicity-density phase diagram for Illustris. The region enclosed by the grey solid line is the metallicity-density space occupied by the gas particles in the Sherwood simulation. The dashed black line is the metallicity-overdensity relation from \citet{keating2014}. Right: mass (dashed) and volume (solid) metallicity filling factors in the four simulations.}
  \label{met_temp_dens}
\end{figure*}

The mean mass-weighted temperature of the gas as a function of overdensity in the simulations is similar in all cases, considering the differences in the hydrodynamic solvers and the different UV backgrounds used (left panel of Figure \ref{met_temp_dens}). The most notable feature is the sharp upturn in the temperature at the lowest overdensities ($\log \Delta \sim -1$), which may correspond to hot gas being pushed out into voids. Alternatively, it may be a hot bubble of gas in pressure equilibrium with colder gas in a higher density environment. The full temperature-overdensity phase diagram for Illustris is also plotted in grey, and it shows that there is a large scatter in the temperature in each overdensity bin. Gas as hot as $T \sim 10^{7}$K is seen, produced by the supernova/AGN feedback and shock heating, but note that this only constitutes a small fraction of all gas.

As one would expect, by increasing the velocity of galactic winds, metals are pushed out to lower density gas (middle panel of Figure \ref{met_temp_dens}). This mass-weighted mean also agrees surprisingly well with the estimate of the metallicity of the $z \sim 6$ CGM by \citet{keating2014}, where the metals were simply added ``by hand'' using a simple metallicity-overdensity relation. The more realistic simulations, however, also predict a large scatter which was neglected in that work. This is shown for Illustris (shaded grey region) and Sherwood (grey outline) which show the scatter in metallicity for each overdensity bin. In the \textsc{arepo} runs, the mixing allows for a large scatter in metallicities, that extends beyond the range shown in the middle panel of Figure \ref{met_temp_dens}. The scatter in metallicity is much smaller in the Sherwood simulations but still spans $\sim$ 5 dex for $\log \Delta > 0$.

We also calculate the filling factor of metals in the simulations using an approach similar to \citet{booth2012}, such that the volume filling factor ($f_{V}$) and the mass filling factor ($f_{m}$) are defined as
\begin{equation}
f_{V} = \frac{\sum_{i} \frac{m_{i}(>Z)}{\rho_{i}(>Z)}}{\sum_{i} \frac{m_{i}}{\rho_{i}}} \, \textnormal{and} \, f_{m} = \frac{\sum_{i} m_{i}(>Z)}{\sum_{i} m_{i}},
\end{equation}
where $i$ is summed over the resolution elements and $Z$ is some threshold metallicity. Depending on the choice of code and feedback model, the volume filling factor of metals can vary by over an order of magnitude (right panel of Figure \ref{met_temp_dens}). Even in the HVEL simulation, only $\sim$ 3 per cent of the volume is enriched to a metallicity $\log (Z/Z_{\odot}) > -6$. In all of the \textsc{arepo} simulations, we find  an (albeit slowly) increasing filling factor over the metallicity range we investigate. If we look down to far lower metallicities, we see filling factors $\sim 1$ due to the possibility for metals to mix. In contrast, the lack of mixing in the SPH simulation is evident from the flattening of both the mass and volume metal filling factors below $\log (Z/Z_{\odot}) \sim -3$. 

\begin{figure*}
\includegraphics[width=2.1\columnwidth]{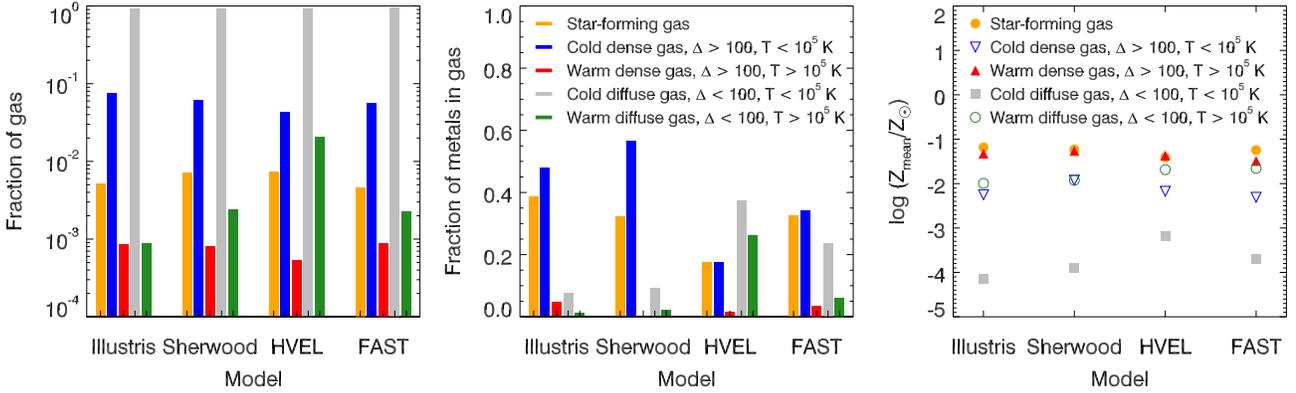}
\caption{Left: The fraction of gas mass in different phases, with cuts made on temperature, overdensity and star-formation rate (the colour scheme is indicated in middle panel) for four different simulations. Middle: The fraction of metal-enriched gas in different phases. Right: The mass-weighted mean metallicity of the gas in each phase.}
  \label{phases}
\end{figure*}

Following the approach of \citet{wiersma2011} and \citet{suresh2015}, we also investigate the metallicity in different gas phases in the simulations. We separate the gas into five categories, based on simple cuts in star-formation, overdensity and temperature:
\begin{itemize}
\item[-] Star-forming gas
\item[-] Cold dense gas, with $\Delta > 100$ and $T < 10^5$ K
\item[-] Warm dense gas, with $\Delta > 100$ and $T > 10^5$ K
\item[-] Cold diffuse gas, with $\Delta < 100$ and $T < 10^5$ K
\item[-] Warm diffuse gas, with $\Delta < 100$ and $T > 10^5$ K
\end{itemize}
The left panel of Figure \ref{phases} shows the fraction of mass of gas in each of these gas phases. The proportion of the gas in each phase is similar across all simulations. The Sherwood and HVEL simulations show slightly more star-forming gas and the HVEL simulation contains the most diffuse gas.

In the middle panel of Figure \ref{phases}, we plot the fraction of mass of the metals in gas in each of these phases. Note that the total mass of metals produced differs only by about a factor of 1.5 across the different simulations. The location of these metals, however, can vary quite considerably. In Sherwood and Illustris, most of the metals (more than 80 per cent) are found in cold dense gas or star-forming gas which, by design, will also be cold and dense. The HVEL and FAST models are successful in pushing a greater proportion of their metals out into lower density gas, as is also shown in the middle panel of Figure \ref{met_temp_dens}. There are some metals in the warm dense gas phase of the Sherwood simulation, but the fraction is very small and is not visible on this plot. The mean mass-weighted metallicity of the gas in each phase is shown in the right panel of Figure \ref{phases}. This shows less variation than the fraction of the metals in each phase. The cold diffuse gas in the HVEL simulation has the highest metallicity compared to the other simulations and it is likely this, together with the fact that the majority of its metals are in diffuse gas, that explains the sharp upturn in metallicity at low gas overdensities in the HVEL simulation seen in the middle panel of Figure \ref{met_temp_dens}.

\subsection{Feedback and Halo Metallicity Profiles}

\begin{figure*}
\includegraphics[width=2.1\columnwidth]{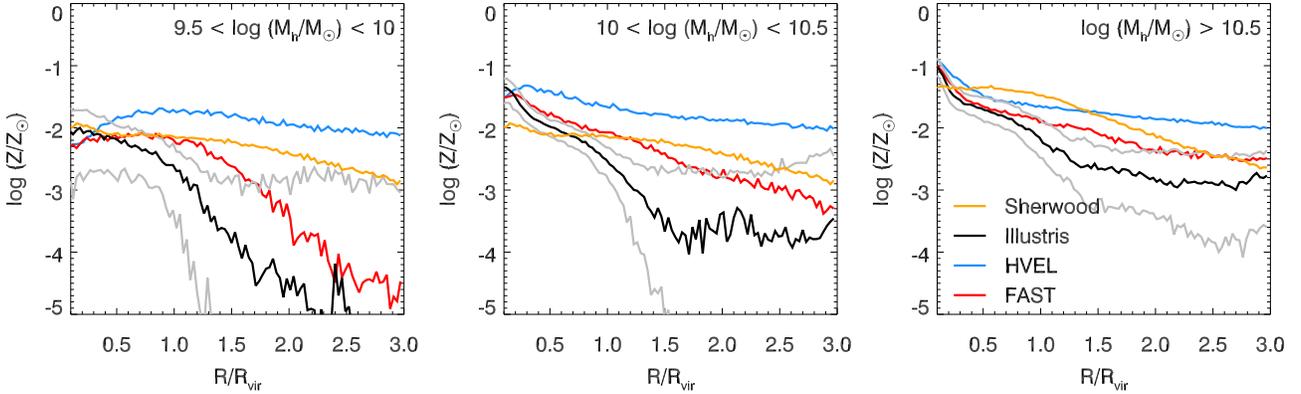}
\caption{The median mass-weighted metallicity around haloes in three different mass bins (left: $9.5 < \log (M_{\textnormal{\scriptsize{halo}}}/M_{\odot}) < 10$, middle: $10 < \log (M_{\textnormal{\scriptsize{halo}}}/M_{\odot}) < 10.5$, right: $\log (M_{\textnormal{\scriptsize{halo}}}/M_{\odot}) > 10.5$) for the four simulations. There are 100 haloes taken from each simulation in every mass bin. The grey lines are the 25$^{\textnormal{\scriptsize{th}}}$/75$^{\textnormal{\scriptsize{th}}}$ percentile scatter in the Illustris simulation.}
  \label{radial_met}
\end{figure*}

We have also investigated how the metallicity profile around haloes changes in the different simulations and as a function of halo mass (shown in Figure \ref{radial_met}). We consider the profiles in three different mass bins and include 100 haloes in each bin. All models are reasonably similar inside 0.5 $R_{\textnormal{\scriptsize{vir}}}$. Beyond that, the models with higher wind speeds transport metals more efficiently away from the hosts. The exception to this is the Sherwood simulation which is surprisingly flat up to 3 $R_{\textnormal{\scriptsize{vir}}}$ in all mass bins. This could be a result of using an SPH code that does not allow for mixing of metals. When we take the median metallicity in each bin over our sample of 100 haloes, we end up with something that looks very flat. It is also worth noting that if our metallicity profiles continued out to larger distances, we would also see the mean metallicity drop to zero at some point as there are many particles in the Sherwood simulation that contain no metals, unlike in the simulations run with \textsc{arepo} (also evident in right panel of Figure \ref{met_temp_dens}).

The trends in the other three runs follow a similar pattern to what is seen in the baryon fractions (right panel of Figure \ref{vel_gasfrac_gsmf}), with the simulations where the metals are pushed out to larger radii corresponding to the ones with lower baryon fractions. The HVEL simulation shows a remarkably flat metallicity across the three bins of halo mass, showing a metallicity $\log (Z/Z_{\odot}) \sim -2$ even at a distance of 3 $R_{\textnormal{\scriptsize{vir}}}$ from the halo. This highlights the effectiveness of this model at enriching the IGM and echoes the flatness of the metallicity-overdensity relation. The Illustris and FAST models also show a shallow metallicity gradient for haloes with $\log (M_{\textnormal{\scriptsize{halo}}}/M_{\odot}) > 10.5$, but the metallicity falls off more steeply as the halo masses decrease (and the wind velocities decrease). Again, these trends reiterate what was already shown in the middle panel of Figure \ref{phases} where we considered the fraction of metals in diffuse gas. 

There is substantial scatter in the metallicity profiles around haloes in the bins with $\log (M_{\textnormal{\scriptsize{halo}}}/M_{\odot}) < 10.5$ (shown here for the Illustris simulation). This is smaller ($\sim 1$ dex) for the haloes with $\log (M_{\textnormal{\scriptsize{halo}}}/M_{\odot}) > 10.5$ but increases hugely beyond the virial radius for less massive haloes. The scatter is most likely driven by other nearby galaxies (also evident in the bumps present in the median value). For the larger galaxies, the metal enrichment will be mostly driven by the strong winds of the massive galaxy and the satellites should not contribute as much.

\section{Metal Absorption Lines}

To make a link between these simulations and observations of metal absorption lines in the spectra of high-redshift QSOs, it is necessary to estimate the fraction of the element in observable ions. Here, we look at three elements (oxygen, carbon and magnesium) and four ions (\ion{O}{i}, \ion{C}{ii}, \ion{C}{iv} and \ion{Mg}{ii}). 

\subsection{Ionization States of Metals}

To estimate the ionic fraction of the metals at a given density and temperature, we used the photo-ionization code \textsc{cloudy} \citep{ferland2013}.   Following the method outlined in \citet{bird2015}, we included a frequency-based attenuation of the UV background in self-shielded regions. We use the overdensity at which an absorber becomes self-shielded \citep[e.g.][]{schaye2001}
\begin{equation}
\label{eqn:ss}
\Delta_{\textnormal{\scriptsize{ss}}} = 54 \, \Gamma^{2/3}_{-12} \, T^{2/15}_{4} \, \left(\frac{1+z}{7}\right)^{-3},
\end{equation} 
which we include in the self-shielding prescription from \citet{rahmati2013}
\begin{equation}
\begin{split}
\frac{\Gamma_{\textnormal{\scriptsize{Phot}}}}{\Gamma_{\textnormal{\scriptsize{HI}}}}
= \, 0.98& \,
\bigg[1+\bigg(\frac{n_{\textnormal{\scriptsize{H,eff}}}}{n_{\textnormal{\scriptsize{H,ss}}}}\bigg)^{1.64}\bigg]^{-2.28}
\\ + \, &0.02 \,
\bigg[1+\frac{n_{\textnormal{\scriptsize{H,eff}}}}{n_{\textnormal{\scriptsize{H,ss}}}}\bigg]^{-0.84}.
\end{split}
\end{equation}
where $n_{\textnormal{\scriptsize{H,eff}}}(E) = n_{\textnormal{\scriptsize{H}}} \, \sigma(1 \, \textnormal{Ryd}) / \sigma(E)$. As $\sigma(E)$ goes like $E^{-3}$, this effect is strongest for the low-energy part of the spectrum and should not affect high-ionization lines like \ion{C}{iv}.

\begin{figure*}
\includegraphics[width=2.1\columnwidth]{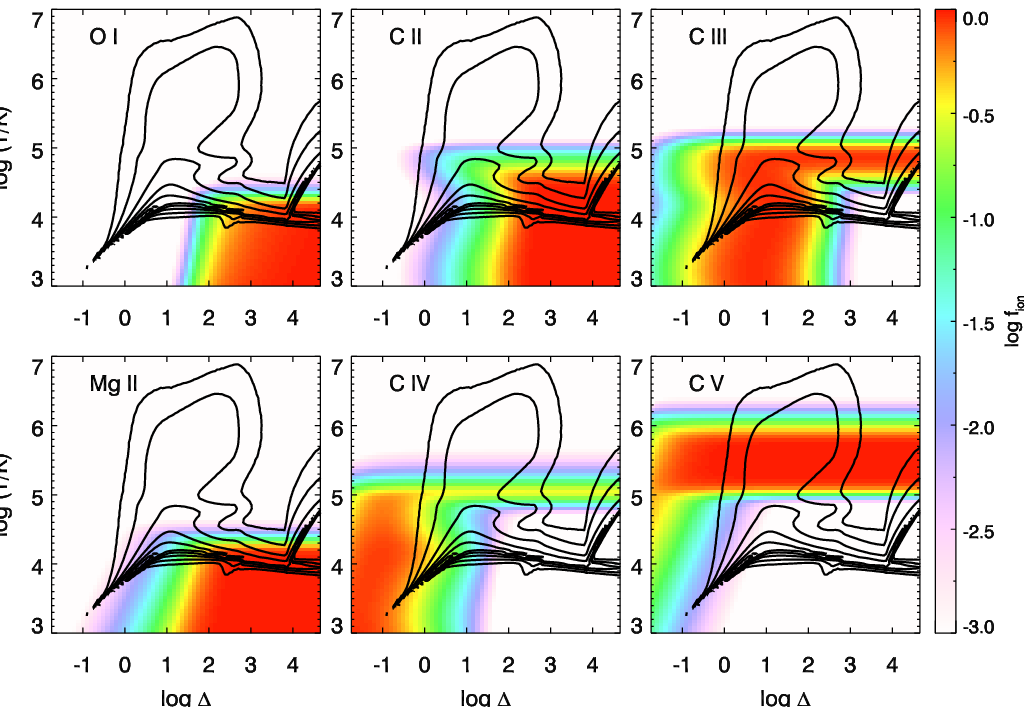}
\caption{The ionic fraction as a function of temperature and density for \ion{O}{i}, \ion{Mg}{ii}, \ion{C}{ii},  \ion{C}{iii}, \ion{C}{iv} and \ion{C}{v} that we calculate from our \textsc{cloudy} models (although note that our models extend beyond the temperature-density range shown here) using the \citet{haardtmadau2012} model of the UV background. Overplotted is the mass-weighted temperature-density distribution in the Sherwood simulation.}
  \label{ionfrac}
\end{figure*}

As expected, we find that the low-ionization ions are associated with dense ($\log \Delta > 1.5$), cold ($\log (T/\textnormal{K}) < 4.5$), self-shielded regions (as shown in Figure \ref{ionfrac}). The \ion{C}{ii} fraction falls off more slowly with decreasing overdensity than \ion{O}{i}, which should result in a larger \ion{C}{ii} covering fraction of haloes and therefore a higher incidence rate than \ion{O}{i} (although this will also depend on the relative abundances of these elements). \ion{Mg}{ii} is found over a similar temperature range to \ion{O}{i}, but extends to lower overdensities. \ion{C}{iv} can be split into two cases: the photoionized \ion{C}{iv}, which peaks at low overdensities ($\log \Delta < 0$) and the collisionally ionized \ion{C}{iv} which peaks at $\log (T/\textnormal{K}) = 5.1$ and is constant across the overdensity range shown here. Part of this is also seen in \ion{C}{ii}, which has an increased ionic fraction at low density gas just above $\log (T/\textnormal{K}) = 4.5$, but the two cases are not as clearly separated as for \ion{C}{iv}. The fraction of collisionally ionized \ion{C}{iv} is at most 30 per cent, with the remainder of the carbon found in \ion{C}{iii} or \ion{C}{v}. \citet{rahmati2015} and \citet{bird2015} have both shown that, at lower redshifts, \ion{C}{iv} is predominantly photoionized. At $z = 6$ however, using the \citet{haardtmadau2012} UV background, most of the photoionized low density gas in the simulations ($0 < \log \Delta < 1$) where \ion{C}{iv} is thought to be found is in the form of \ion{C}{iii}. We discuss how this changes for different UV backgrounds in Section \ref{sec:shapeUV}. 

\begin{figure*}
\includegraphics[width=2.1\columnwidth]{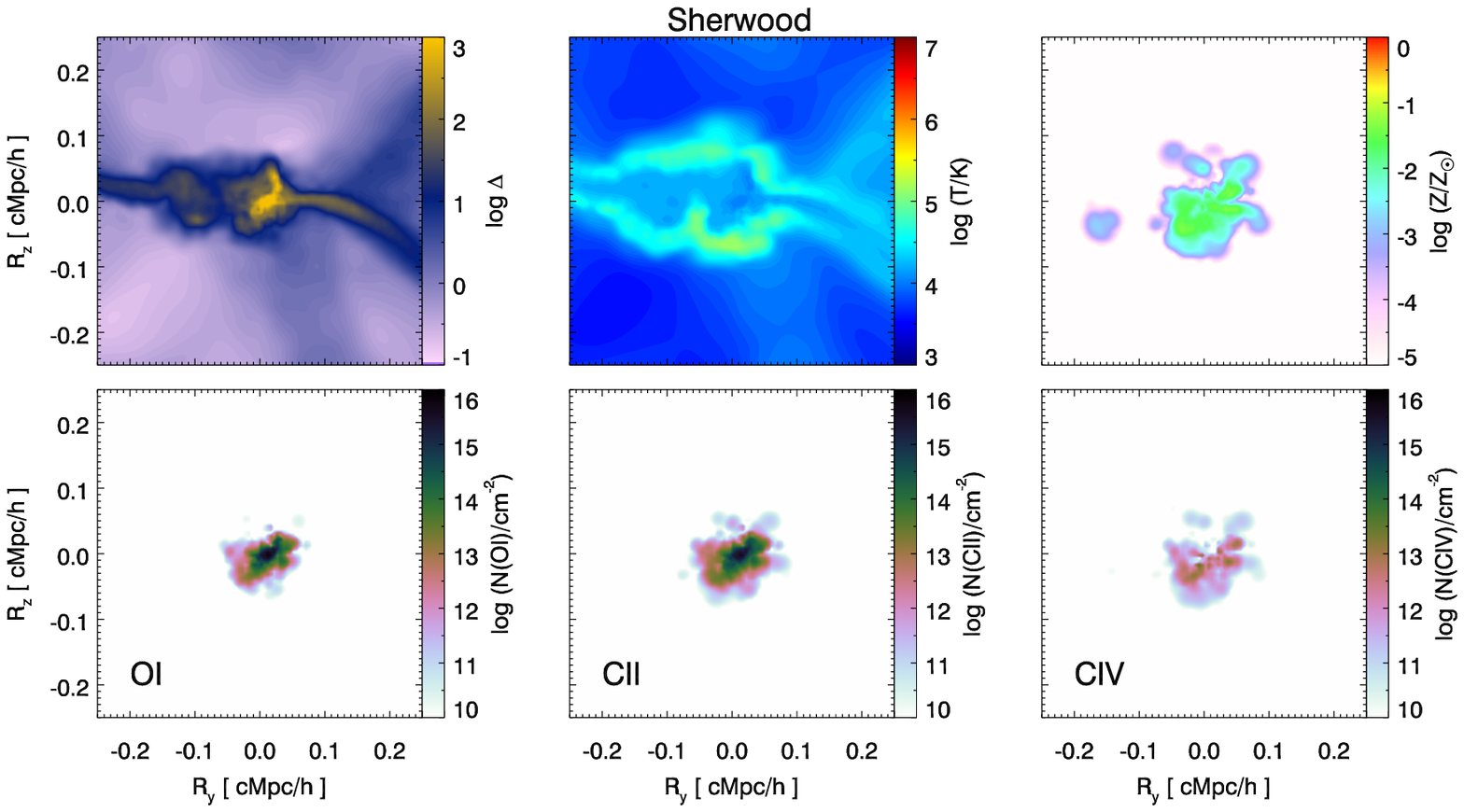}
\caption{Slices around a halo with $M_{\textnormal{\scriptsize{halo}}} = 10^{10} \, M_{\odot}$ from the Sherwood simulation. The top panel shows the overdensity (left), the temperature (middle) and the metallicity (right). The bottom panel shows the column densities of \ion{O}{i} (left), \ion{C}{ii} (middle) and \ion{C}{iv} (right). The thickness of the slice is 4.9 ckpc $h^{-1}$.}
  \label{prace_maps}
\end{figure*}

\begin{figure*}
\includegraphics[width=2.1\columnwidth]{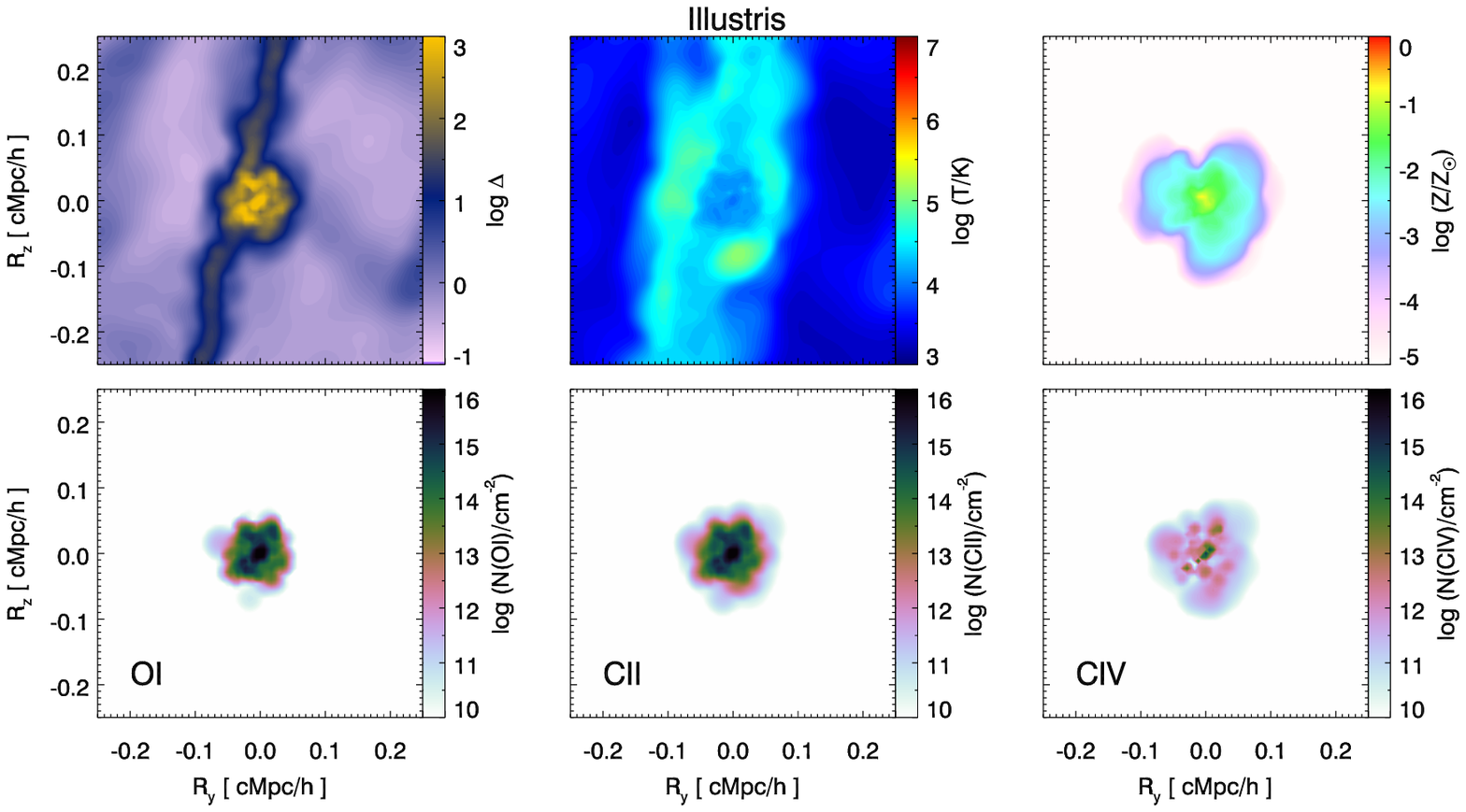}
\caption{Slices around a halo with $M_{\textnormal{\scriptsize{halo}}} = 10^{10} \, M_{\odot}$ from the Illustris simulation. The top panel shows the overdensity (left), the temperature (middle) and the metallicity (right). The bottom panel shows the column densities of \ion{O}{i} (left), \ion{C}{ii} (middle) and \ion{C}{iv} (right). The thickness of the slice is 5.7 ckpc $h^{-1}$.}
  \label{ill_maps}
\end{figure*}

We next used these models to assign ionic fractions to the particles/resolution elements in our simulations. Examples of what this looks like for a $10^{10} \, M_{\odot}$ halo in the Sherwood and Illustris simulations are shown in Figures \ref{prace_maps} and \ref{ill_maps}. We have tried to select two haloes whose density fields look similar, however keep in mind that these haloes are different and therefore we will only make a qualitative comparison. In both cases, the size of the metallicity bubble (where the metallicity is $\log (Z/Z_{\odot}) > -4$) is comparable to the region occupied by hot gas. In the \textsc{arepo} simulations, the metallicity continues to fall off slowly but it becomes so small that it will not produce observable absorbers and is therefore not plotted here. The distribution of the metals appears somewhat clumpier in the Sherwood simulation, possibly because the metals are not able to mix in this case.

In the bottom panel of both figures, we plot the column densities for three of the ions we are interested in: \ion{O}{i}, \ion{C}{ii} and \ion{C}{iv}. The low-ionization absorbers fill only a small part of the metallicity bubble and are split into small clumps, which track the position of the cold, metal-enriched gas. The size of the clumps seems to be larger in Illustris, maybe because the gas density is higher and the metallicity is higher, but it is not clear if this is because we are looking at two different haloes or because of differences in hydro-solvers in \textsc{gadget-3} \textit{vs.} \textsc{arepo}. In both cases, the \ion{C}{iv} extends over a larger area than the low-ionization absorbers. The larger covering fraction is in agreement with the higher incidence rate observed for \ion{C}{iv} absorbers by \citet{dodorico2013} than for the incidence rate of low-ionization absorbers by \citet{becker2011}. The strongest column density \ion{C}{iv} absorbers are still found in the centre of the halo, however, as this is where the metallicity is highest. 

\begin{figure*}
\includegraphics[width=2.1\columnwidth]{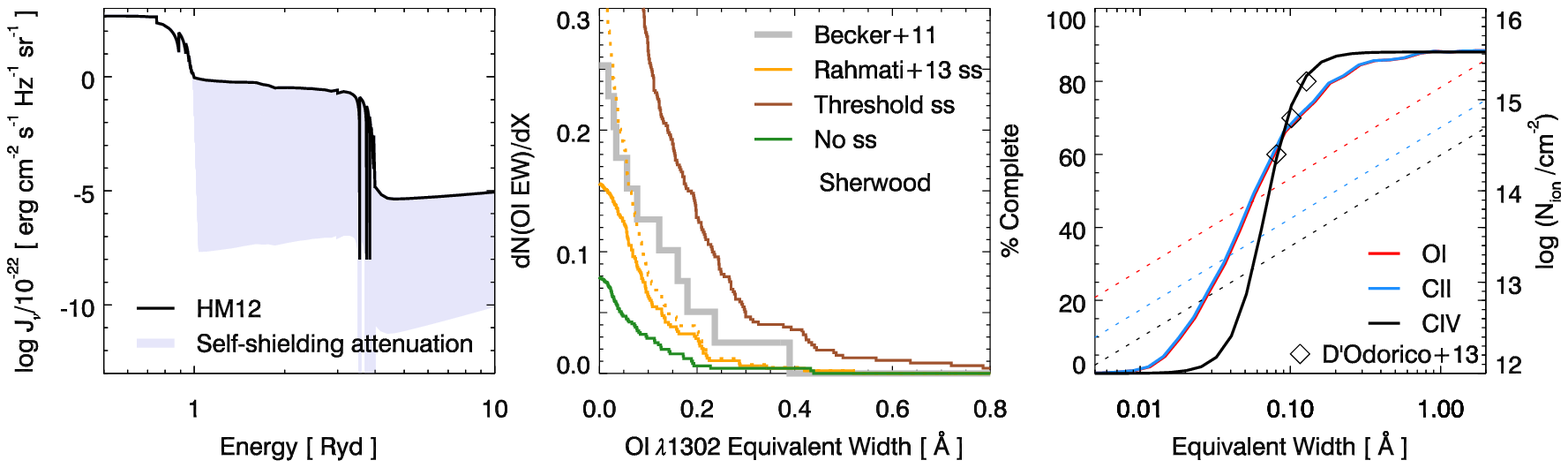}
\caption{Left: the \citet{haardtmadau2012} model for the UV background at $z=6$, shown in black. The shaded region shows the effect of the frequency based self-shielding, with the lowest intensity corresponding to a model with $T = 10^{3}$ K and $n_{\textnormal{\scriptsize{H}}} = 10^{4}$ cm$^{-3}$. Middle: distribution of \ion{O}{i} equivalent widths, assuming no self-shielding, a model based on the \citet{rahmati2013} prescription and with a simple threshold self-shielding model. Shown in grey are the observational data from \citet{becker2011}. The solid lines are the completeness-corrected distributions and the dashed line is the full distribution we model. Right: The percentage completeness of equivalent width (solid lines) we assume for  \ion{O}{i}, \ion{C}{ii} and \ion{C}{iv}. The open diamonds are the estimate for the \ion{C}{iv} completeness from \citet{dodorico2013}. The dashed lines show the relation we assume between column density and equivalent width.}
  \label{uvgamma}
\end{figure*}

\subsection{Synthetic Spectra}

\begin{table}
\centering
\begin{tabular}{c|c|c|c|c}
Ion & $\lambda$ (\AA) & $f$ & $\Delta E$ (eV) & $\log Z_{\odot}$  \\
\hline
\ion{O}{i} & 1302.2 & 0.05 & $<$ 13.62  & -3.31 \\
\ion{C}{ii} & 1334.5 & 0.13 & 11.26 -- 24.38& -3.57 \\
\ion{C}{iv} & 1548.2 & 0.19& 47.89 -- 64.49 & -3.57 \\
\ion{Mg}{ii} & 2796.4 & 0.63& 7.64 -- 15.03 & -4.40 \\
\end{tabular}
\caption{Atomic parameters of the ions we focus on in this paper: \ion{O}{i}, \ion{C}{ii}, \ion{C}{iv} and \ion{Mg}{ii}. Shown are the wavelength ($\lambda$), the oscillator strength ($f$), the energy required to bring the element into this ionic state and the energy that will further ionize it ($\Delta E$) and the metal abundances we assume ($\log Z_{\odot}$), taken from \citet{asplund2009}.}
\label{ions}
\end{table}

\begin{figure*}
\includegraphics[width=2.1\columnwidth]{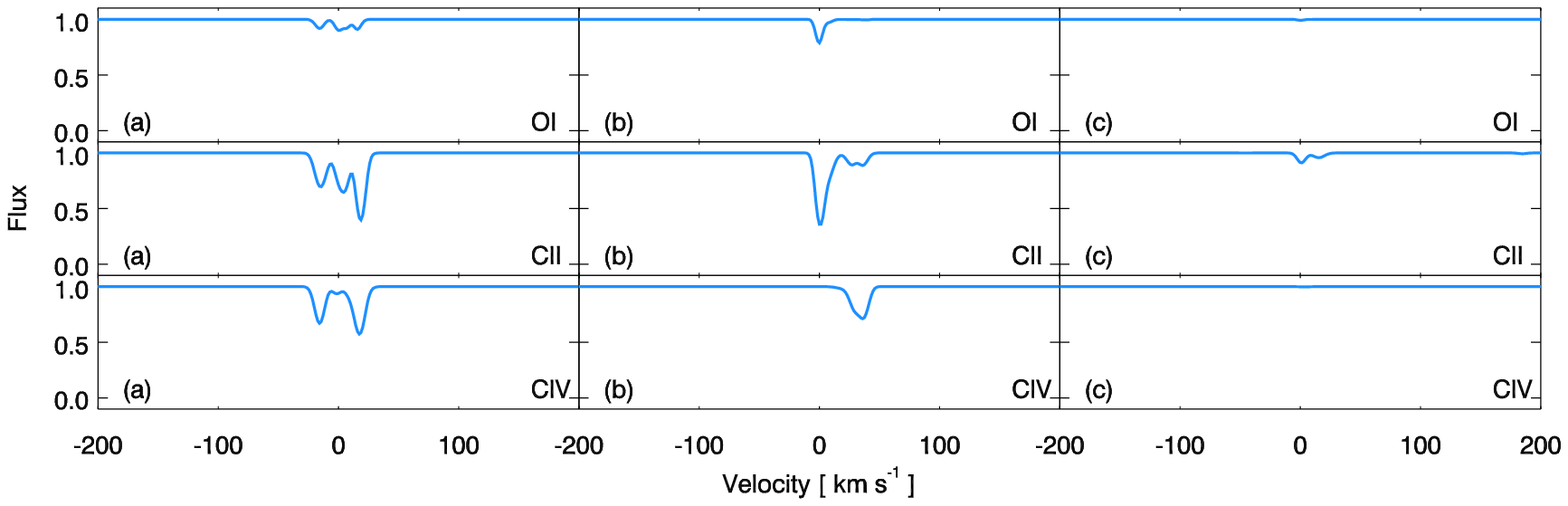}
\caption{Mock spectra generated along three sightlines (shown in the left, middle and right columns) taken from the Sherwood simulation. The first row shows the \ion{O}{i} spectra, the middle shows \ion{C}{ii} and the bottom row shows \ion{C}{iv}. We define $v = 0$ km s$^{-1}$ by the deepest part of the \ion{O}{i} absorption for each sightline.}
\label{spectra}
\end{figure*}

We construct artificial spectra along random sightlines through the simulation volume using the ionic density together with the ion-weighted temperature and peculiar velocity at each pixel. The relevant atomic parameters are summarised in Table \ref{ions}. For \ion{C}{iv} and \ion{Mg}{ii}, we only use the lines at 1548 \AA \, and 2796 \AA \, respectively, neglecting the other half of the doublets. Figure \ref{spectra} shows examples of spectra along three of our sightlines. The first sightline displays an absorption feature that is generally the same shape for all three ions. The depth of the absorption varies however, due to differences in the oscillator strength of that ion and also in the fraction of that element that is in this state. The second sightline now shows how different ions can trace different kinematic features - there is a velocity offset between the \ion{O}{i} and the \ion{C}{iv}, while the \ion{C}{ii} traces both components. Finally, in the third sightline, we see a \ion{C}{ii} line only, perhaps because the gas is too dense to see \ion{C}{iv} and the weak oscillator strength of \ion{O}{i} prevents a clear absorption feature. 
 
From these spectra, we measure the equivalent width of the absorption systems and then apply a completeness correction to make a comparison with observations. In the case where there are two or more absorption features within 50 km s$^{-1}$ of each other, then we treat these as one system with multiple components. However, our results were not very sensitive to this number and we also produced similar equivalent width distributions assuming absorption features within 200 km s$^{-1}$ of each other were one system. For \ion{O}{i} and \ion{C}{ii}, we use an estimate for completeness taken from \citet{becker2011}. For \ion{C}{iv}, we fit a curve of the form
\begin{equation}
f(x) = \frac{L}{1 + \mathrm e^{-k(x-x_0)}},
\end{equation}
to the estimates for completeness from \citet{dodorico2013}, i.e. that the data are (60,70,85) per cent complete for column densities $\log (N_{\textnormal{\scriptsize{\ion{C}{iv}}}}/\textnormal{cm}^{-2}) = (13.3,13.4,13.5)$. We take $(L,k,x_0) = (88.1,9.1,13.2)$. These completeness estimates are shown in the middle pane of Figure \ref{uvgamma}. We also plot the analytic relation between equivalent width (EW) and column density ($N$) in the regime where optical depths are small:
\begin{equation}
\textnormal{EW} = N\frac{ \pi q_{e}^{2} \lambda^{2} f}{m_{e} c^2},
\end{equation}
where $q_{e}$ is the electron charge, $\lambda$ is the transition wavelength, $f$ is the oscillator strength and $m_{e}$ is the electron mass. 

To determine the frequency of our simulated absorbers, we normalise the total number of absorbers by the pathlength $\Delta X$ corresponding to the number of sightlines times the size of our box, where $X$ is defined as 
\begin{equation}
\label{eq:pathlength}
X(z) = \int^z_0 \, \frac{H_0}{H(z')} (1+z')^2 \, \rmn{d}z'.
\end{equation}
We show the cumulative distribution of \ion{O}{i} absorbers we obtain in the right panel of Figure \ref{uvgamma} for the Sherwood simulation. We show here three different cases - one where we include no self-shielding, one with the \citet{rahmati2013} self-shielding prescription and finally one with a simple threshold self-shielding model, where all gas above a self-shielding overdensity is considered to be neutral.  The effect of including the \citet{rahmati2013} self-shielding prescription on the shape of the UV background is shown in the left panel of Figure \ref{uvgamma}, where the most extreme attenuation shown corresponds to $T = 10^{3}$ K and $n_{\textnormal{\scriptsize{H}}} = 10^{4}$ cm$^{-3}$. As expected, the models without self-shielding under-predict the incidence rate of absorbers, especially the weaker ones. Using the threshold self-shielding model leads to an incidence rate that is far in excess of observations. The \citet{rahmati2013} model seems to agree reasonably well with observations (a more detailed comparison with the data is made in Section \ref{sec:comparison}), and this is the approach we take in all the work that follows. The dashed line shows the absorber distribution we would obtain without applying a completeness correction. As expected from the right panel of Figure \ref{uvgamma}, this makes little difference to absorbers with equivalent width greater than 0.1 \AA. Below this, we see a rapid increase in the incidence rate of absorbers that are generally below the detection limit of current observations. 

\section{Comparison with Observations}
\label{sec:comparison}

\begin{figure*}
\includegraphics[width=2.1\columnwidth]{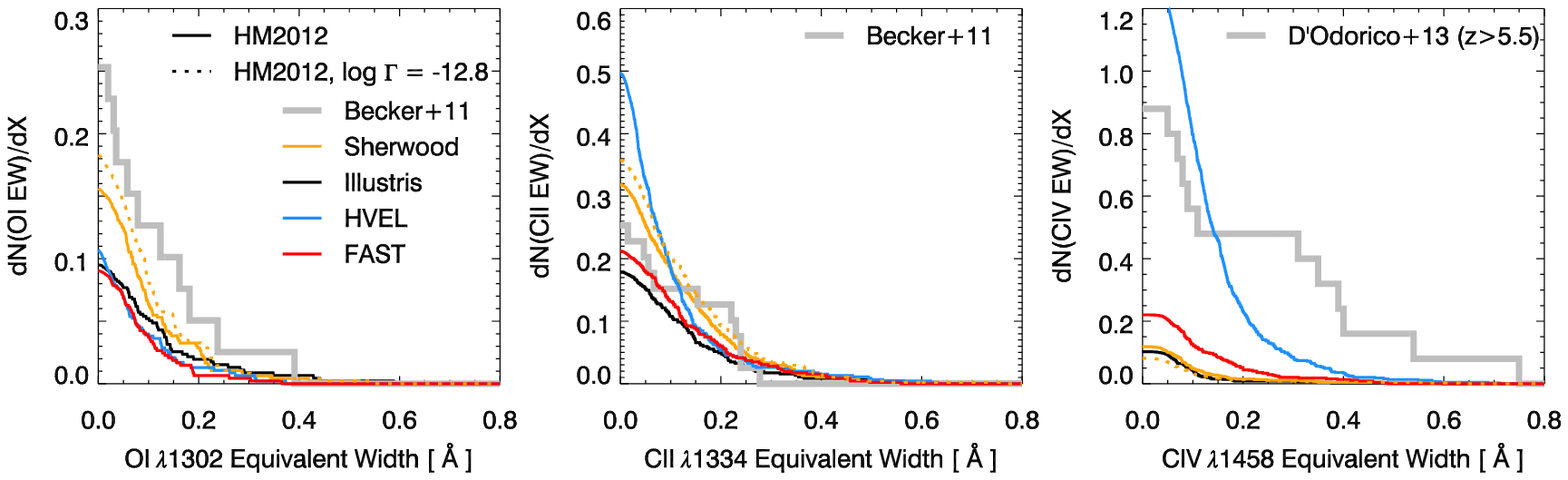}
\caption{Cumulative distribution of equivalent widths for \ion{O}{i} (left), \ion{C}{ii} (middle) and \ion{C}{iv} (right) in different simulations. The grey lines are the data from \citet{becker2011} and \citet{dodorico2013}. A correction to match the completeness of the observations has been applied to the simulated absorbers, which has the effect of removing some weak absorbers that may not be detected in the observations. The solid line is for ion densities calculated using the \citet{haardtmadau2012} UV background and the dashed line shows the a UV background with the same shape but with its amplitude rescaled to give a background photoionization rate $\log (\Gamma_{\textnormal{\scriptsize{\ion{H}{i}}}}/\textnormal{s}^{-1}) = -12.8$, in line with observations at $z \sim 6$.}
  \label{ew_feedback}
\end{figure*}

\begin{figure*}
\includegraphics[width=2.1\columnwidth]{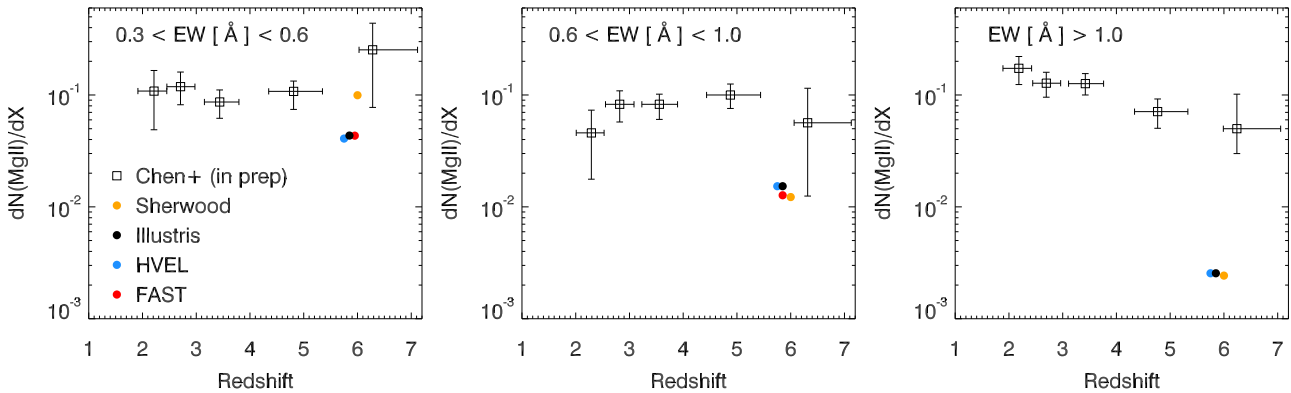}
\caption{The \ion{Mg}{ii} incidence rate in three separate equivalent width bins: 0.3 $< \textnormal{EW}_{\textnormal{\scriptsize{\ion{Mg}{ii}}}} \, [ \textnormal{\AA} ] <$ 0.6 (left panel), 0.6 $< \textnormal{EW}_{\textnormal{\scriptsize{\ion{Mg}{ii}}}} \, [ \textnormal{\AA} ] < $1.0 (middle panel) and $\textnormal{EW}_{\textnormal{\scriptsize{\ion{Mg}{ii}}}} \, [ \textnormal{\AA} ] >$ 1.0 (right panel), using the \citet{haardtmadau2012} UV background. The observed data is taken from Chen et al (in prep) and is based on work by \citet{matejek2012}. No completeness correction was applied here. Note that there were no absorbers with $\textnormal{EW}_{\textnormal{\scriptsize{\ion{Mg}{ii}}}} \, [ \textnormal{\AA} ] >$ 1.0 in the FAST model. We have adjusted the redshifts slightly so that all points are visible.}
  \label{mgii_dndx}
\end{figure*}

We choose to compare our mock spectra with the real observations by looking at the distribution of equivalent widths of our absorbers. Although this can introduce some uncertainty due to the line shapes induced by kinematics of the gas, using the equivalent widths allows us access to a slightly larger sample of low-ionization absorbers as \citet{becker2011} only compute column densities for their high-resolution sample of spectra (seven out of ten absorbers). 

\subsection{Feedback Models and Metal Line Absorbers}
\label{sec:feedback_ew}

In Figure \ref{ew_feedback}, we show the cumulative distribution functions of the equivalent width observations of high-redshift \ion{O}{i}, \ion{C}{ii} and \ion{C}{iv} absorbers. All our models of \ion{O}{i} absorbers have incidence rates below the \citet{becker2011} line-of-sight number density, but almost match the lower 95 percent confidence interval (which is based on Poisson errors only). We under predict the abundance of weak \ion{O}{i} absorbers, but this could be explained by uncertainty in the completeness of the data, in the metal yields or in the treatment of self-shielding.  Increasing the resolution of the simulations also increases the number of weak absorbers produced. This is discussed further in Appendix \ref{sec:restest}. Alternatively, it may also be indicative of a lower background photoionization rate at $z=6$. We also plot the \citet{haardtmadau2012} UV background rescaled to give a background photoionization rate $\log (\Gamma_{\textnormal{\scriptsize{\ion{H}{i}}}}/\textnormal{s}^{-1}) = -12.8$, motivated by \citet{calverley2011} and \citet{wyithe2011}. This is equivalent to rescaling the amplitude of the UV background to 0.6 of the original amplitude. This results in an increased number of weak absorbers (EW$_{\textnormal{\scriptsize{\ion{O}{i}}}} < 0.2 \, \textnormal{\AA}$), while not adding many strong ones. The overall shape of our \ion{O}{i} equivalent width distribution is similar to the observed distribution. Better agreement for the stronger absorbers could therefore be reached by increasing the metallicity by a factor of a few, which is likely reasonable within the given uncertainties in metal yields and relative elemental abundances of these absorbers. Alternatively, more optimistic assumptions about the amount of self-shielded, neutral gas present would certainly increase the incidence rate of strong absorbers (see, for example, the middle panel of Figure \ref{uvgamma}). There does also appear to be some dependence on box size, with the stronger absorbers found in larger volumes. We also match the velocity width distribution of the absorbers quite well, which is discussed in more detail in Appendix \ref{sec:velwidths}.

\begin{figure*}
\includegraphics[width=2\columnwidth]{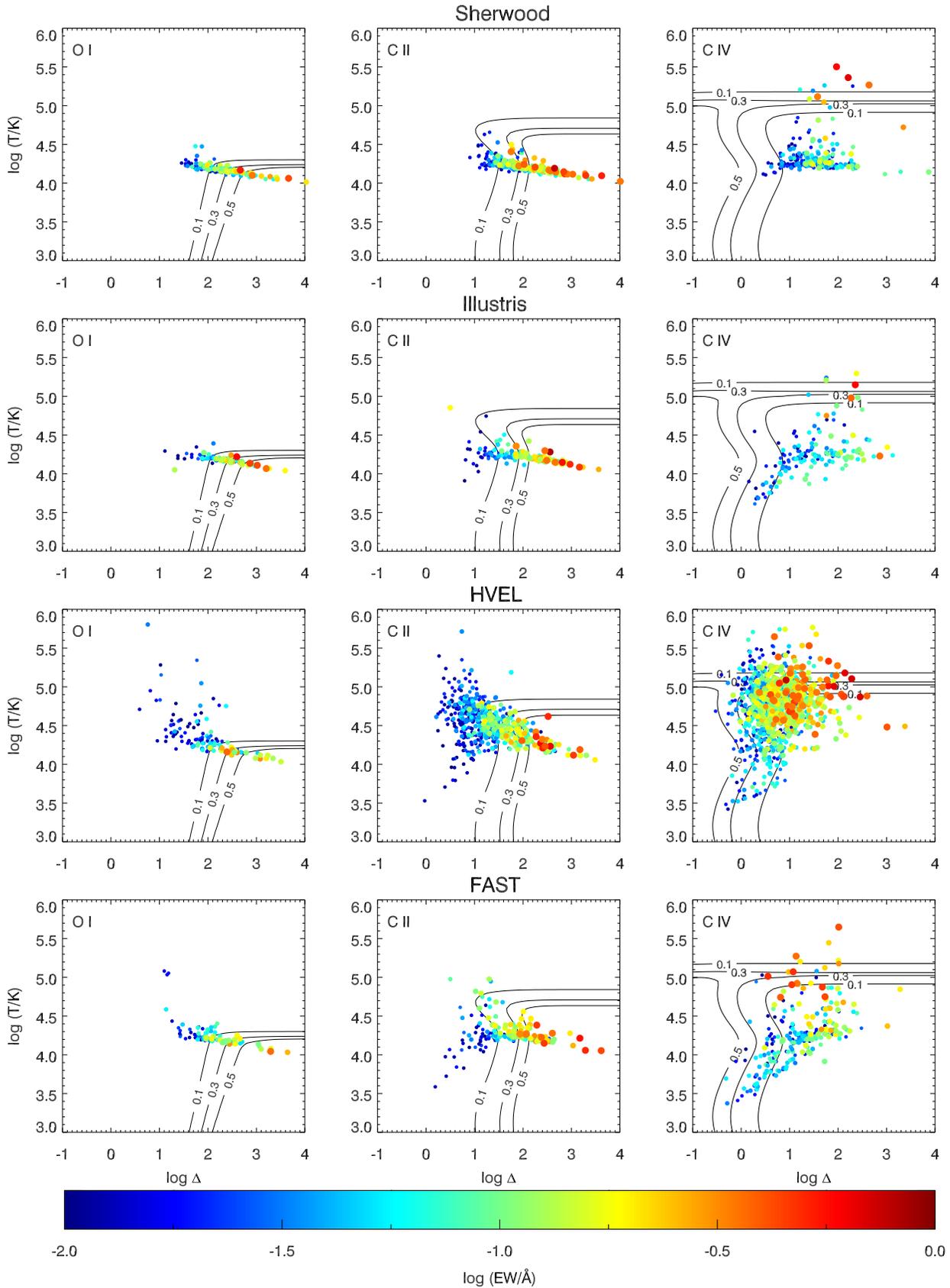}
\caption{Ion-weighted temperature-overdensity plots for the \ion{O}{i}, \ion{C}{ii} and \ion{C}{iv}  absorbers we measure in the four simulations. The colour of the points corresponds to the equivalent width and the black contours enclose the regions where the ionic fraction is 10, 30 and 50 per cent.}
  \label{ew_temp_dens}
\end{figure*}

The general agreement between the four models can be explained by the similar mean metallicity at the overdensities $\Delta \gtrsim 100$ where a significant fraction of \ion{O}{i} is found (see middle panel of Figure \ref{met_temp_dens} and Figure \ref{ionfrac}). The story is similar for \ion{C}{ii}, with three of our models providing a decent match to the observations. However, the HVEL model hugely over-predicts the number of weak absorbers, as the fall-off in the metallicity-overdensity relationship is much shallower for this model and \ion{C}{ii} probes out to lower overdensities than \ion{O}{i}. The temperature of these absorbers is also higher ($\sim 10^{4.5}$ K), which allows for higher \ion{C}{ii} fractions at lower overdensities (Figure \ref{ionfrac}). It is also worth noting that the simulations run with \textsc{arepo} generally predict a larger number of weak absorbers (Figure \ref{ew_temp_dens}), due to the mixing of metals down to lower density gas. For the purposes of this work, this is inconsequential as they lie below the detection threshold of current instruments. It may however become more important in the future, when making predictions for next-generation telescopes. We also note that although we find a higher incidence rate of \ion{C}{ii} here compared with \ion{O}{i}, which is in line with what is expected from the ionization fractions calculated in our \textsc{cloudy} models, this will be dependent on the relative abundances of carbon and oxygen. Here, we assume solar abundances but the value at high-redshift may be lower \citep{becker2012}, which would perhaps result in a similar incidence rate for detectable \ion{C}{ii} and \ion{O}{i} absorbers. 

We find that matching the observations of \ion{C}{iv} and \ion{Mg}{ii} absorbers, however, is much more challenging. The Sherwood, Illustris and FAST models under-predict the observed incidence rate of \ion{C}{iv} absorbers by a factor of 5--10. The HVEL model manages to reproduce the total incidence rate, but the shape of the cumulative distribution function does not match the data, with not enough strong absorbers (EW$_{\textnormal{\scriptsize{\ion{C}{iv}}}} \gtrsim 0.2 \, \textnormal{\AA}$) seen. Using the model with a lower background photoionization rate reduces the incidence rate of weak \ion{C}{iv} absorbers further. For \ion{Mg}{ii}, we match observations of absorbers with 0.3 $\, <$ EW$_{\textnormal{\scriptsize{\ion{Mg}{ii}}}} \, [ \textnormal{\AA} ] <$ 0.6 (left and middle panels of Figure \ref{mgii_dndx}) reasonably well. However, we have not applied a completeness correction here so the agreement would become more tenuous if this were taken into account. We also struggle to match the incidence rate for absorbers with $\textnormal{EW}_{\textnormal{\scriptsize{\ion{Mg}{ii}}}} \, [ \textnormal{\AA} ] >$ 1.0, falling far short of the observations at $z=6$ and with the FAST simulation failing to produce any absorbers of this strength at all. The main difference between the low- and high-ionization absorbers is the overdensity at which the ionic fraction peaks. Therefore, the issues with modelling the  \ion{C}{iv} and \ion{Mg}{ii} absorbers are most likely related to how efficiently the low overdensities are enriched with metals. For the rest of this paper, we focus on why we struggle to match the observations of \ion{C}{iv} absorbers but note that reproducing the incidence rate of \ion{Mg}{ii} absorbers is most likely a related problem.

In Figure \ref{ew_temp_dens}, we show how the equivalent width correlates with the temperature and overdensity of these absorbers in each of our simulations. Plotted also are contours showing where the fraction of that ion is 10, 30 or 50 per cent. We find that the strongest absorbers (EW $\gtrsim 0.1$ \AA) tend to be found at overdensities $\log \Delta > 1.5$ as this is where the metallicity is highest. More low-density absorbers are seen in the FAST and HVEL models, due to the increase in the amount of metal-enriched, low-density gas in these simulations. The \ion{C}{iv} absorbers tend to be found over a wider temperature range ($T = 10^{3.5-5.5}$ K) than the low-ionization absorbers and also probe out to lower densities $\log \Delta \sim 0$. However, the peak of the photoionized \ion{C}{iv} fraction occurs below the overdensities probed here which may prevent us from seeing stronger \ion{C}{iv} absorbers. Indeed, most of the strong \ion{C}{iv} absorbers we see tend to be collisionally ionized. This is discussed further in Section \ref{sec:shapeUV}. \citet{dodorico2013} used \textsc{cloudy} models to suggest that their \ion{C}{iv} absorbers were found in gas with $\delta \sim 10$ (where $\delta = \Delta - 1$), but only the HVEL and FAST models have enough metals at this overdensity to produce any strong absorbers at $\delta \sim 10$. 

\begin{figure*}
\includegraphics[width=2.1\columnwidth]{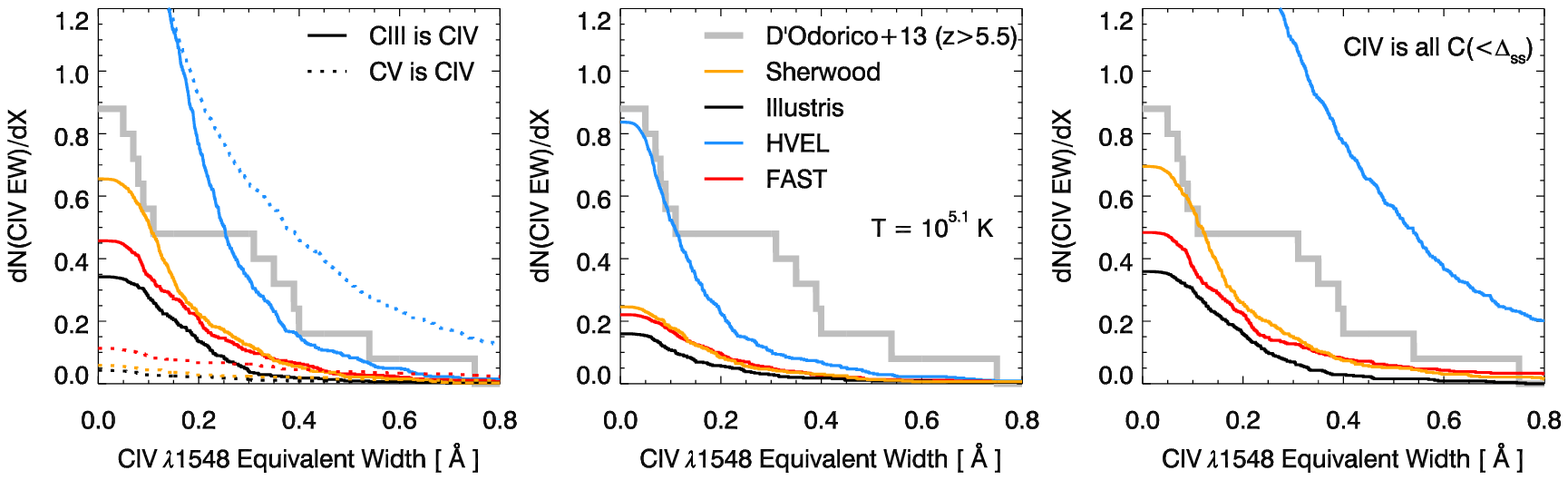}
\caption{Left: Cumulative distribution of equivalent widths of \ion{C}{iv} absorbers in different simulations, checking what happens if we take the \ion{C}{iii} (solid line) or \ion{C}{v} (dashed line) fraction instead of \ion{C}{iv}. The grey lines are the data from \citet{dodorico2013}. Middle: distribution of \ion{C}{iv} absorbers if the gas is all at $10^{5.1}$ K. In these cases we start to find some stronger absorbers.  Right: Cumulative distribution of \ion{C}{iv} equivalent widths, now assuming that all of the carbon is \ion{C}{iv} (unless the gas is self-shielded). Even for this somewhat unrealistic assumption, the Sherwood, Illustris and FAST models only just about reach the incidence rate of \ion{C}{iv} absorbers. A correction to match the completeness of the observations has been applied to the simulated absorbers.}
  \label{civ_ion_temp}
\end{figure*}

At this point, the main difficulty with modelling metal-line absorbers becomes clear: is the lack of \ion{C}{iv} absorbers a problem with the enrichment model or with the ionization state of our simulations? We try to understand this further in Figure \ref{civ_ion_temp}. Here we look at the \ion{C}{iv} absorption we would see if instead of taking the fraction of carbon in \ion{C}{iv}, we take the fraction in \ion{C}{iii} or \ion{C}{v}. We find that a lot of the carbon that could produce the weak absorption systems  (EW$_{\textnormal{\scriptsize{\ion{C}{iv}}}} \lesssim 0.2 \, \textnormal{\AA}$) is in the form of \ion{C}{iii}. This could potentially be moved into \ion{C}{iv} with a harder UV background. We also find that the HVEL model would overproduce the incidence rate of \ion{C}{iv} absorbers if the gas that was in \ion{C}{v} was instead found in \ion{C}{iv} (perhaps if it was not shock-heated by the fast winds, for example). The Sherwood, Illustris and FAST simulations still fail to produce many \ion{C}{iv} absorbers suggesting that shock-heated gas is not an issue for these models. We also check how our distribution of \ion{C}{iv} absorbers would look if all the gas was at $T =  10^{5.1}$ K, where the collisionally ionized \ion{C}{iv} fraction peaks. The Sherwood, Illustris and FAST models still fail to reproduce the observed incidence rate however and the HVEL model results in far too many weak absorbers. This result can be understood by looking at Figure \ref{ionfrac}. Even at $T =  10^{5.1}$ K, \ion{C}{iv} makes up only 30 per cent of all the carbon, with most of the carbon is found in \ion{C}{iii} and \ion{C}{v}.

As a simple test, we also tried assuming that all carbon below the threshold overdensity where self-shielding becomes important was \ion{C}{iv} (right panel of Figure \ref{civ_ion_temp}). We take this threshold overdensity from the relation in Equation \ref{eqn:ss}, which gives $\Delta_{\textnormal{\scriptsize{ss}}} = 21.7$ at $T = 10^{4}$ K. This scenario is obviously somewhat unrealistic, as from Figure \ref{ionfrac} we see that even if the gas was all at $T =10^{5.1}$ K, only about 30 per cent would be collisionally ionized into \ion{C}{iv}. Even in this toy model, however, the Sherwood, Illustris and FAST simulations only just about reach the observed incidence rate of  \ion{C}{iv} absorbers. This suggests that the metals are not being pushed out to low enough densities in these models, and we will always find it hard to match the $z \sim 6$ \ion{C}{iv} incidence rate, regardless of the choice of UV background or the temperature of the gas. In this case, the HVEL model does seem to enrich the IGM sufficiently, however this model is disfavoured by models of DLAs at $z = 2-4$ \citep{bird2014,bird2015} and would almost certainly fail to reproduce the shape of the galactic stellar mass function at $z=0$.

\subsection{Shape of the UV Background}
\label{sec:shapeUV}

\begin{figure*}
\includegraphics[width=2.1\columnwidth]{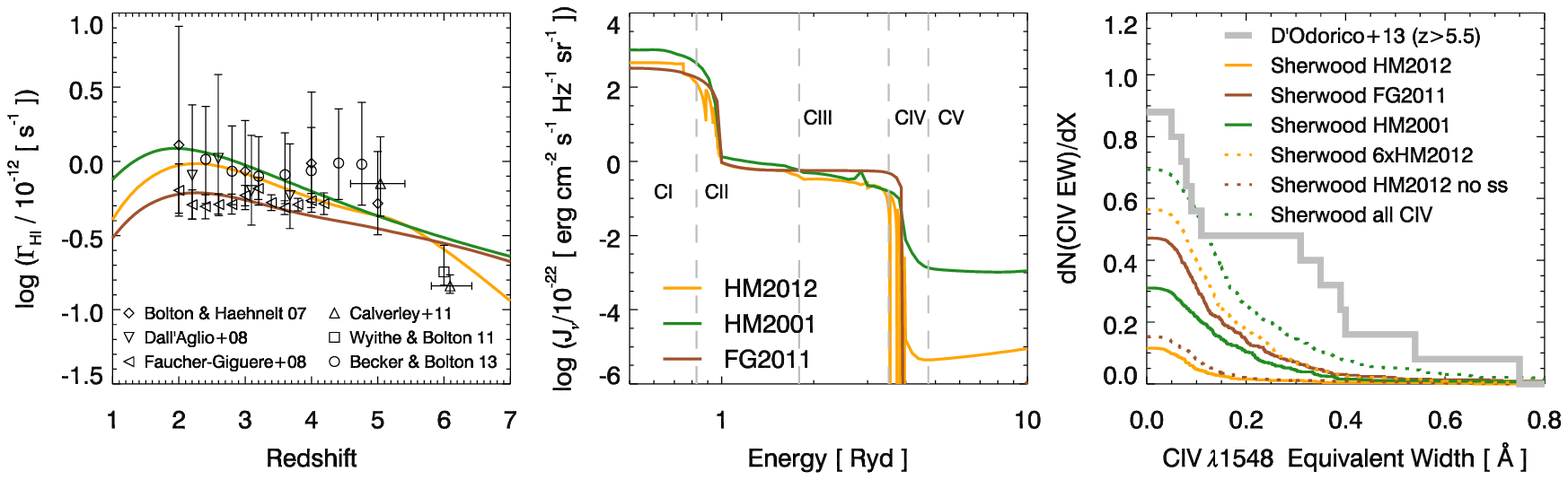}
\caption{Left: Evolution of the background photoionization rate $\Gamma_{\textnormal{\scriptsize{\ion{H}{i}}}}$ with redshift in the \citet{haardtmadau2012}, the 2011 update of \citet{fauchergiguere2009} and \citet{haardtmadau2001} models. The colour scheme is the same as the middle panel. For comparison we show observations from \citet{bolton2007gammahi}, \citet{dallaglio2008}, \citet{fauchergiguere2008}, \citet{calverley2011}, \citet{wyithe2011} and \citet{becker2013}. Middle: Shape of the UV background as a function of energy for three different models at $z=6$. Also shown are the relevant carbon ionization energies. Right: Cumulative distribution of \ion{C}{iv} equivalent widths for the Sherwood simulation, for UV backgrounds with different shapes and amplitudes, a model without self-shielding and a model assuming all the non-self-shielded carbon was \ion{C}{iv}. Changing the UV background increases the number of weak absorbers but fails to produce many stronger ones.}
  \label{uv_civ}
\end{figure*}

\begin{figure*}
\includegraphics[width=2.1\columnwidth]{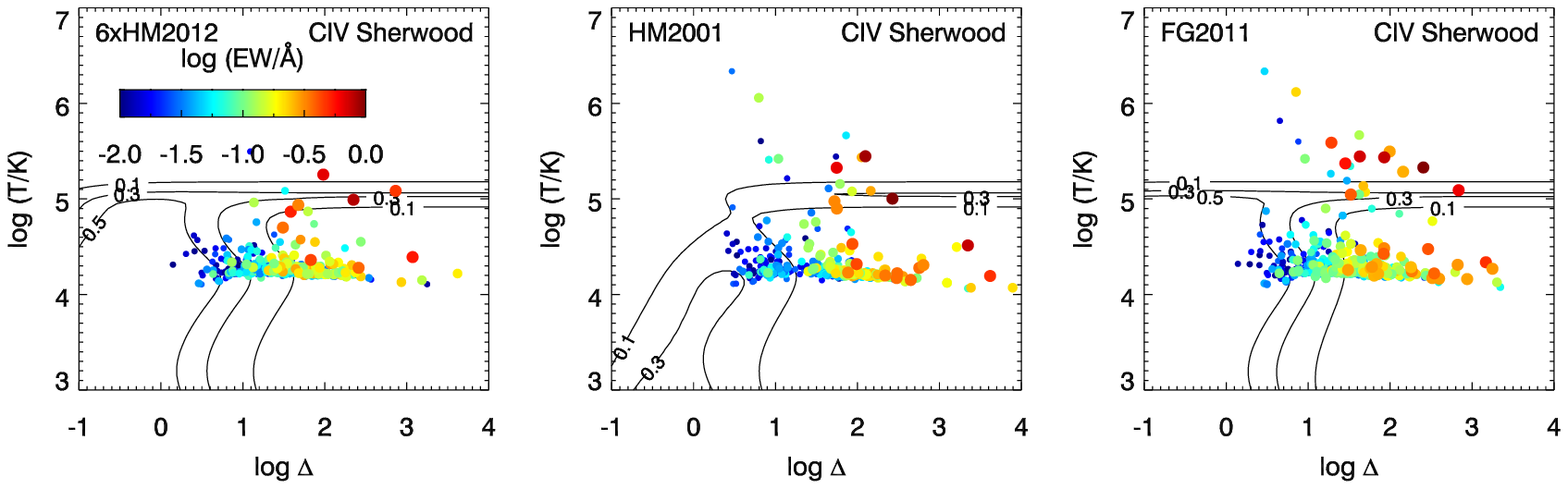}
\caption{\ion{C}{iv}-weighted temperature-overdensity plots for the \ion{C}{iv} absorbers we measure in the Sherwood simulation using three different UV backgrounds. Left: \citet{haardtmadau2012} a factor six higher in amplitude, middle: \citet{haardtmadau2001}, right: the 2011 version of \citet{fauchergiguere2009}. The colour of the points corresponds to the equivalent width, with the colour bar in the left panel showing the values for all three plots. The black contours enclose the temperatures and densities for which the fraction of carbon in \ion{C}{iv} is 10, 30 and 50 per cent.}
  \label{uv_civ_temp_dens}
\end{figure*}

Even though Section \ref{sec:feedback_ew} leads us to conclude that it is the metal enrichment rather than the ionization model that results in our dearth of \ion{C}{iv} absorbers, it is still worth investigating how the uncertainty in the shape and amplitude of the UV background at $z \sim 6$ would affect our results. We note also that, although we do not consider it here, looking at the ratio of column densities of different ions in absorbers (e.g. $\log \, N_{\textnormal{\scriptsize{\ion{C}{iv}}}}$ \textit{vs.}  $\log \, N_{\textnormal{\scriptsize{\ion{S}{iv}}}}$) is a powerful way of placing constraints on the shape of the UV background.  We explore the effect of changing the UV background using the Sherwood simulation in the right panel of Figure \ref{uv_civ}. We first confirm that our self-shielding prescription does not have a significant effect on the \ion{C}{iv} absorption, as one would expect for a high-ionization ion.

We next investigate how the \ion{C}{iv} absorbers are affected by changing the shape of our input UV background. This is somewhat motivated by observations of hard spectra in some high-redshift galaxies, such as \citet{stark2015} who found \ion{C}{iv} emission in a $z \sim 7$ Ly$\alpha$ emitter which was comparable to that of an AGN. There has also been recent discussion on the contribution of quasars at $z \sim 6$ \citep{chardin2015,haardt2015,madau2015}, driven by updated quasar luminosity functions \citep{giallongo2015} and new constraints on the optical depth to reionization \citep{planck2015}. We explore the effect of using two other commonly implemented UV backgrounds: \citet{haardtmadau2001} and the 2011 update of \citet{fauchergiguere2009}, as well as increasing the amplitude of the \citet{haardtmadau2012} background by a factor of six to mimic a locally enhanced amplitude due to nearby star-formation and/or spatial UV fluctuations at the tail-end of reionization. We show the \ion{H}{i} photoionization rates predicted by these models as a function of redshift in the right panel of Figure \ref{uv_civ}. For comparison we also show measurements of the photo-ionization rate out to $z \sim 6$, which shows that all of the models are at the high end of the error range given by the \citet{wyithe2011} result. This also suggests that increasing the amplitude of the \citet{haardtmadau2012} model by a factor of six would be incompatible with observations, however in reality there will be fluctuations in the photoionization rate depending on proximity to galaxies, etc. \citep[e.g.,][]{becker2015,chardin2015,finlator2015}.  We show the intensity as a function of energy for the different models in the middle panels of Figure \ref{uv_civ}. The ionization energies for the different states of carbon (up to \ion{C}{v}) are also shown. In the energy range where \ion{C}{iv} is dominant, the three models have very different shapes due to the treatment of the sources and the absorbers (here, most importantly the helium absorbers). Examples of the fraction of each ion seen at different densities and temperatures at each of these UV backgrounds are shown in Appendix \ref{sec:ions_uvbs}.

We find that the  \citet{haardtmadau2001}, \citet{fauchergiguere2009} and 6$\times$ \citet{haardtmadau2012} models all result in an enhancement in the \ion{C}{iv} absorbers with EW$_{\textnormal{\scriptsize{\ion{C}{iv}}}} \lesssim 0.4 \, \textnormal{\AA}$ (right panel of Figure \ref{uv_civ}). The different models still (unsurprisingly) fall short of our toy model where all the carbon is \ion{C}{iv} and we still fail to produce the strongest observed \ion{C}{iv} absorbers, but it is evident that changing the UV background does push things in the right direction. This is further shown in Figure \ref{uv_civ_temp_dens}, where the temperature and overdensity of the \ion{C}{iv} absorbers is plotted together with the \ion{C}{iv} fraction from our \textsc{cloudy} models. The three models allow for higher \ion{C}{iv} fractions at higher overdensities than with the \citet{haardtmadau2012} UV background. The result of this is an overall increase in the total number of absorbers, more absorbers at high overdensities ($\log \Delta > 2$) and stronger absorbers at moderate overdensities ($0.5 < \log \Delta < 1.5$) than when compared with the top right panel of Figure \ref{civ_ion_temp}. Most of the strong \ion{C}{iv} absorbers, however, still occur at temperatures $T > 10^{5}$ K, indicating that they are collisionally ionized.

\subsection{Absorber-Galaxy Connection}

\begin{figure*}
\includegraphics[width=2\columnwidth]{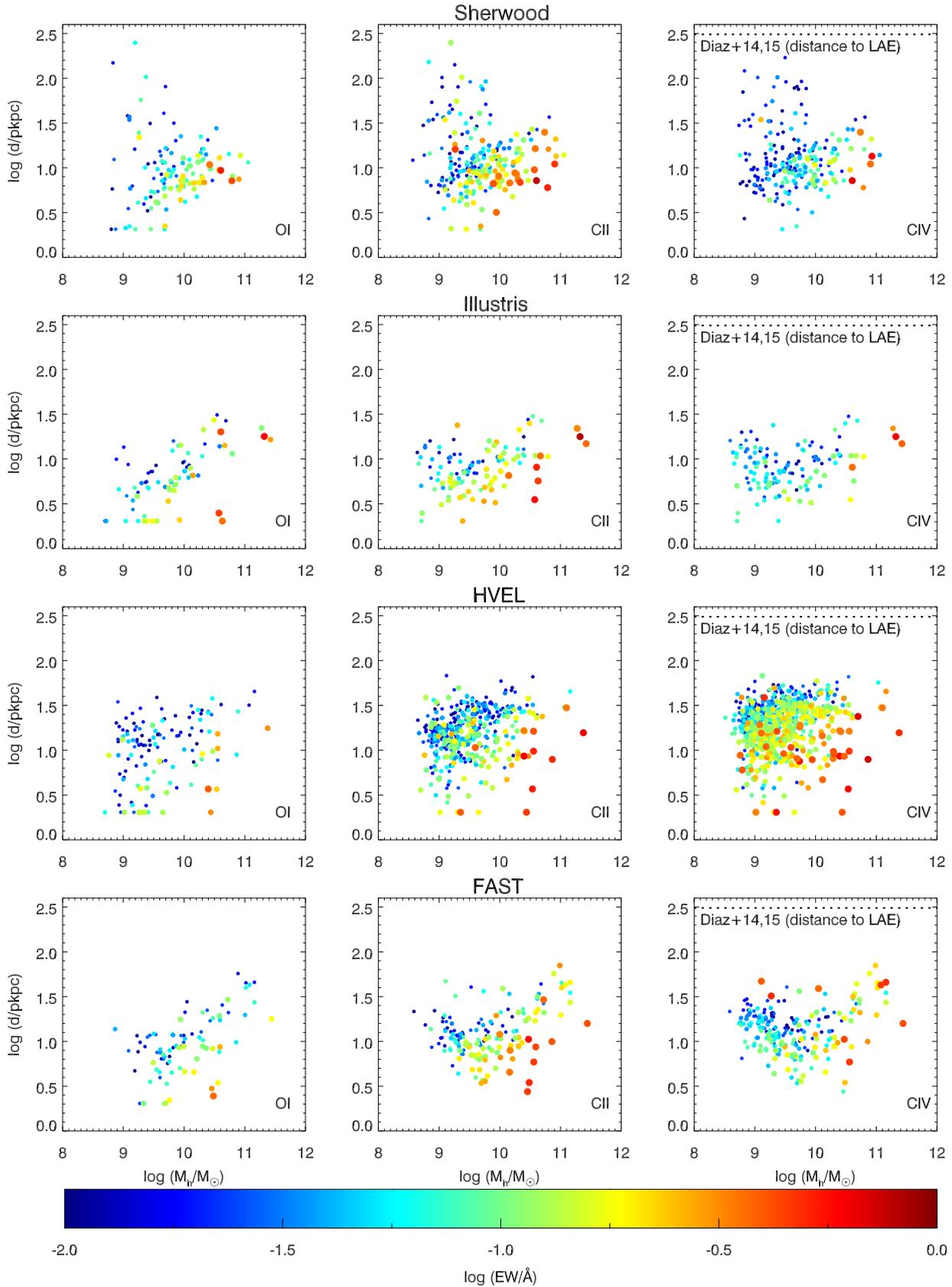}
\caption{The mass of the halo associated with a metal-line absorber against the distance from that absorber in the four simulations. The points are coloured by the equivalent width of the absorber and the colour scheme is the same across all panels. Also shown is the distance between a $z \sim 5.7$ \ion{C}{iv} absorber and the nearest spectroscopically confirmed Ly$\alpha$ emitter \citep{diaz2014,diaz2015}.}
  \label{halo_abs}
\end{figure*}

It is also interesting to look into the connection between metal-line absorbers and galaxies. These were selected by making an association between the absorber with the largest equivalent width along a sightline and the nearest halo. The position of the absorber was determined by taking the pixel with the maximum ion number density, weighted by the total ion number density along the sightline. We constrained the halo to have a non-zero stellar mass and a total mass $\log (M_{\textnormal{\scriptsize{halo}}}/M_{\odot}) > 8.5$ to avoid selecting small satellite haloes. The results are shown in Figure \ref{halo_abs}. We find that the strongest absorbers are located closest to the host haloes and the haloes tend to have masses $\log (M_{\textnormal{\scriptsize{halo}}}/M_{\odot}) \gtrsim 10$. As the mass of the host halo increases, strong absorbers can be found at increasing distances from the halo. Occasionally we find strong absorbers linked to haloes below $10^{10} \, M_{\odot}$, but this may be due to an absorber be linked to a satellite halo of a nearby, more massive halo. Across all the simulations, the low-ionization lines tend to be strongest when the absorbers are found in close proximity to a halo. In the HVEL simulation, strong \ion{C}{iv} absorbers tend to be found closer to haloes than in Illustris and FAST. This is likely due to the increased amount of shock-heated gas around these haloes, due to the strong winds, that allows for collisionally ionized \ion{C}{iv} to be present. 

Our results are generally in good agreement with those of \citet{oppenheimer2009}, in the case where they use a uniform UV background. Our haloes span a similar range of stellar masses in the range $6 < \log (M_{*}/M_{\odot}) < 9.5$, apart from in the Sherwood simulation which contains haloes with lower stellar masses due to its higher resolution. The absorbers associated with these low stellar mass haloes are all very weak though ($\textnormal{EW} \lesssim 0.05$ \AA, suggesting that increased resolution will only produce more weak absorbers. The HVEL simulation is the one most like the \citet{oppenheimer2009} constant UV background model in terms of the maximum distance we find absorbers at. However, they find some strong \ion{C}{iv} absorbers at $\log (d/\textnormal{pkpc}) > 1.5$, which we do not see here.

Also marked on the right column of Figure \ref{halo_abs} is the measurement of the distance between a $z \sim 5.7$ \ion{C}{iv} absorption system and a Ly$\alpha$ emitter by \citet{diaz2015}. This absorber, first discovered by \citet{ryanweber2009} was also part of the \citet{dodorico2013} sample who measured an rest-frame equivalent width of 0.75 \AA. They measure that \ion{C}{iv} absorber is 212.8 $h^{-1}$ pkpc from the Ly$\alpha$ emitter. This is further than any of the distances we measure between absorbers and haloes in our simulations. They suggest it is also possible that an undetected dwarf galaxy may be responsible for the metal enrichment, but we find that absorbers with large equivalent widths are generally hosted by haloes with $\log (M_{\textnormal{\scriptsize{halo}}}/M_{\odot}) \gtrsim 10$. However, as none of our models can reproduce the observed incidence rate of \ion{C}{iv} absorbers, it is difficult to say more about this.

\subsection{Evolution of the Mass Density with Redshift}

\begin{figure*}
\includegraphics[width=2.1\columnwidth]{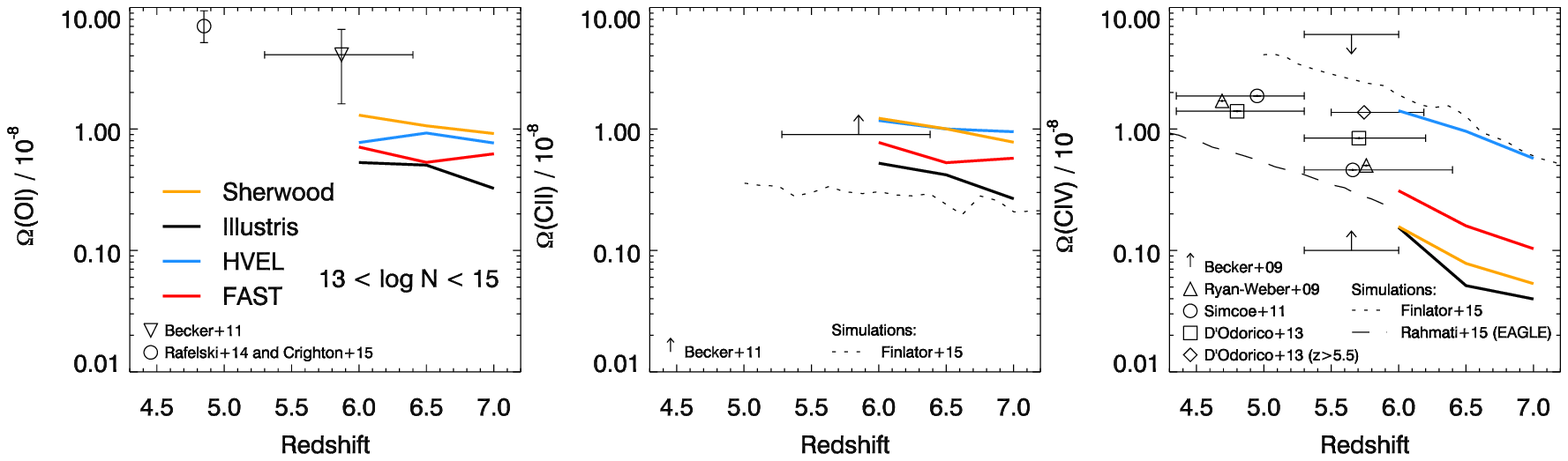}
\caption{Evolution of the mass density of \ion{O}{i} (left), \ion{C}{ii} (middle) and \ion{C}{iv} (right) with redshift. Plotted for comparison are mass densities calculated from observed absorption systems. For \ion{O}{i} we show an estimate for $\Omega_{\textnormal{\scriptsize{\ion{O}{i}}}}$ at $z \sim 5$, taken from \citet{becker2015rev} and calculated from measurements of  $\Omega_{\textnormal{\scriptsize{\ion{H}{i}}}}$ in DLAs \citep{crighton2015} and the mean metallicity of DLAs at that redshift \citep{rafelski2014}. The other points are taken from observations by \citet{becker2011}, \citet{becker2009}, \citet{ryanweber2009}, \citet{simcoe2011} and \citet{dodorico2013}. We also show a point showing only the $z > 5.5$ absorbers in the \citet{dodorico2013} sample. We also compare with calculated mass densities from simulations by \citet{finlator2015} and the EAGLE simulation \citep{rahmati2015}.}
  \label{civ_obs}
\end{figure*}
Finally we look at how the mass density of \ion{O}{i}, \ion{C}{ii} and  \ion{C}{iv} absorbers evolve towards $z=7$. Since we do not have a model that reproduces the observations of \ion{C}{iv} at $z \sim 6$, we do not intend to make predictions for the exact values of  $\Omega_{\textnormal{\scriptsize{\ion{C}{iv}}}}$ that may be observed, but instead just explore the general trends we find. The mass density of an ion is given by
\begin{equation}
\Omega_{\textnormal{\scriptsize{ion}}} = \frac{H_{0} m_{\textnormal{\scriptsize{ion}}}}{c \rho_{\textnormal{\scriptsize{crit}}}} \int^{N_{\textnormal{\scriptsize{max}}}}_{N_{\textnormal{\scriptsize{min}}}}N_{\textnormal{\scriptsize{ion}}}f(N_{\textnormal{\scriptsize{ion}}},z)\textnormal{d}N_{\textnormal{\scriptsize{ion}}} ,
\end{equation}
where $\Omega_{\textnormal{\scriptsize{ion}}}$ is the mass density of a given ion, $m_{\textnormal{\scriptsize{ion}}}$ is the mass of the ion, $c$ is the speed of light, $\rho_{\textnormal{\scriptsize{crit}}}$ is the critical density at $z=0$ and $N_{\textnormal{\scriptsize{ion}}}$ is the column density of the absorber. $f$ is the column density distribution function defined as
\begin{equation}
f(N_{\textnormal{\scriptsize{ion}}},z) = \frac{\textnormal{d}^{2}n}{\textnormal{d}X\textnormal{d}N_{\textnormal{\scriptsize{ion}}}} ,
\end{equation}
where $n$ is the number of absorbers and $X$ is the pathlength defined in Equation \ref{eq:pathlength}. This can be reduced to
\begin{equation}
\Omega_{\textnormal{\scriptsize{ion}}} = \frac{H_{0} m_{\textnormal{\scriptsize{ion}}}}{c \rho_{\textnormal{\scriptsize{crit}}}}  \frac{\sum N_{\textnormal{\scriptsize{ion}}}}{\Delta X} ,
\end{equation}
where the sum is across the detected absorbers. We measure the column density across the absorber by summing the calculated column density at each pixel from its optical depth, using the apparent optical depth method as in \citet{becker2011}, where 
\begin{equation}
N_{\textnormal{\scriptsize{ion}}} = \frac{3.768 \times 10^{14}}{f \lambda}  \sum_{i}\tau_{i}\delta v_{i} \, \textnormal{cm}^{-2}.
\end{equation}
Here $\tau$ is the optical depth, $\delta v$ is the width of the pixel in km s$^{-1}$ and $i$ sums over the pixels associated with the absorber. We checked that assigning a maximum value to the optical depth of $\tau = 5$ made no difference to our results. The mass densities we obtained are presented in Figure \ref{civ_obs} and are shown for the \citet{haardtmadau2012} UV background. 

To try and make a fair comparison with observations, we restrict our measurement of $\Omega_{\textnormal{\scriptsize{ion}}}$ to include only absorbers with column densities in the range $13 < \log (N_{\textnormal{\scriptsize{ion}}}/\textnormal{cm}^{-2}) < 15$ which is in reasonable agreement with the column density range probed by observations. We do not apply a completeness correction on top of this cut, to make comparison with published results from other simulations. Note that this may result in a misleading comparison with observed estimates, which use only the observed (incomplete) set of column densities. The mass density of the low-ionization absorbers are quite sensitive to the range of column densities considered, with the mass density of \ion{O}{i} and \ion{C}{ii} increasing by a factor of a few if we include all absorbers. It is the contribution of the strongest absorbers that matters for this, with the inclusion of absorbers with low column densities having little effect. Although these high-column density systems are rare, they contain a lot of mass which can alter the result significantly. This is also the reason that the \ion{C}{iv} is insensitive to the range of column densities included: our models do not produce the strong absorbers in this case. 

The evolution of \ion{O}{i} and \ion{C}{ii} is reasonably flat between $z = 6$ and 7. Our simulated \ion{O}{i} absorbers fall somewhat short of the measurement of \citet{becker2011}, but again this could be adjusted by assuming a UV background corresponding to a lower photoionization rate or a slightly higher metallicity in the simulations. The \ion{C}{ii} absorbers just about match the lower limit provided by the observations of \citet{becker2011}. We do not find the increase in the incidence rate of \ion{O}{i} absorbers from $z= 6-7$ predicted by \citet{keating2014}, due to the inhomogeneous distribution of metals occupying only a small filling factor of the box in the simulations considered here. 

As with the equivalent width distributions, our models of \ion{C}{iv} again struggle to match the measurements of the \ion{C}{iv} mass density calculated from observations. Similar to what \citet{bird2015b} found at $z =4$, the \ion{C}{iv} mass density calculated from Illustris falls short of that calculated from observations, but the increased velocity FAST model improves things somewhat. However, at $z=6$, this improvement is not enough to provide a convincing match to the observed data. Although the HVEL model appears to come close to matching the observed quantities, it is important to note that it does not reproduce the observed distribution of column densities, with most of the absorbers in the HVEL model having column densities less than $\log (N_{\textnormal{\scriptsize{\ion{C}{iv}}}}/\textnormal{cm}^{-2}) \sim 14$. We also emphasise that we are not applying a completeness correction here, only a cut in column density, which would reduce the mass density in the HVEL model as the observations are only $\sim 20$ per cent complete at $\log (N_{\textnormal{\scriptsize{\ion{C}{iv}}}}/\textnormal{cm}^{-2}) = 13$.

\subsection{Comparison with Other Work and Discussion}

\citet{finlator2015} have modelled the abundance of \ion{C}{ii} and \ion{C}{iv} absorbers during the reionization epoch, at $z > 5$. They use a ``hybrid'' wind model, energy-driven for galaxy velocity dispersions $\sigma < 75$ km s$^{-1}$ and momentum-driven otherwise. A variation of this model, which was momentum-driven only, was shown in \citet{oppenheimer2006} to reproduce $\Omega_{\textnormal{\scriptsize{\ion{C}{iv}}}}$ over the redshift range $2 < z < 6$. The novel aspect of \citet{finlator2015} was to use a coupled radiative transfer and hydrodynamic code, which is certainly a more realistic approach than using a uniform UV background. The downside to this, however, is that the simulations become very expensive and therefore only a 6 $h^{-1}$ Mpc volume was simulated. They find that their simulated UVB results in a \ion{C}{ii} (\ion{C}{iv}) fraction a factor of two lower (higher) from $5 < z < 10$ than the ion fractions calculated with a uniform UV background. This would improve our results but, as shown in the left panel of Figure \ref{uv_civ}, we would probably still not have enough \ion{C}{iv} to match the observations. The middle and right panels of Figure \ref{civ_obs} show that we indeed produce more \ion{C}{ii} and less \ion{C}{iv}, which is in agreement with what is predicted by their simulated UVB but may also point to a difference between the wind models. It seems that the wind model used in \citet{finlator2015} and associated works is very efficient at enriching the IGM with metals, but it is not entirely clear why. Comparing their wind prescription with the Illustris wind physics suggests that, for haloes of the same mass, their winds should have lower velocities and lower mass-loading factors. There is, however, a difference in the density threshold for star formation. Compared to the Sherwood simulation, where the hydrodynamics and resolution should be similar, they use a lower threshold density ($n_{\textnormal{\scriptsize{H}}}$ = 0.13 cm$^{-3}$ \textit{vs.} $n_{\textnormal{\scriptsize{H}}}$ = 0.33 cm$^{-3}$). Note that in both simulations wind particles are decoupled from the hydrodynamics when they enter the wind and are recoupled when the ambient density falls below 10 per cent of the star formation threshold. Thus, the recoupling happens at lower density in their runs, which will most likely facilitate the escape of winds to large radii. Further differences could be due to our inclusion of a completeness correction or the metallicity of the winds (the metallicity in the Illustris winds is only 40 per cent of the ISM metallicity). The justification for this is that \citet{vogelsberger2013} find that winds with the metallicity of the ISM lead to an overenrichment of gas in the haloes of low-mass galaxies. However, setting the metallicity of the winds to 70 or 100 per cent of the ISM metallicity instead was found to have little effect by \citet{bird2015} and \citet{suresh2015}, respectively.

\citet{rahmati2015} have looked at the mass density of \ion{C}{iv} in the EAGLE simulation at $z \sim 6$, which uses a feedback scheme in which no velocity or mass-loading factor is explicitly specified for the galactic winds. Although the main focus of the paper is on highly-ionized metals at lower redshifts, it seems from their mass density of \ion{C}{iv} that they are also struggling to match the observed \ion{C}{iv} abundance at $z \sim 6$ (right panel of Figure \ref{civ_obs}). Note that their models were run using the \citet{haardtmadau2001} model for the UV background, which we find produces more \ion{C}{iv} than the  \citet{haardtmadau2012} model at $z \sim 6$ (Section \ref{sec:shapeUV}).

Our results suggest that a high wind velocity is needed to pollute the low-density regions which are probed by \ion{C}{iv} with metals. Given constraints on the available energy and momentum for driving the winds, this translates to low wind mass-loading factors. However, models with such low mass-loadings typically overpredict the high-redshift star formation rate density as well as the faint end of the galaxy luminosity function \citep[e.g.][]{oppenheimer2006,Puchwein2013gsmf}. It is, thus, not obvious how to resolve both problems at the same time. Improved agreement with the data could probably be achieved by more optimistic assumptions about the available energy and momentum for driving winds, which would allow higher mass loading factors for fast winds. Another route might be to facilitate the escape of the winds from galaxies. For the simple wind models considered here, this could in principle be achieved by increasing the period for which the wind particles are decoupled from the hydrodynamics after entering the wind, which is supposed to mimic the formation of chimneys of relatively low resistance where outflows might escape more easily. Admittedly, increasing the domain of application of what is essentially a sub-resolution treatment of the outflow physics in such a way is not entirely satisfactory. A more gratifying approach might be to follow the physics of outflows in more detail, i.e. at higher resolution and including more of the relevant physics, like cosmic rays that might continue to accelerate the winds on their way out. In addition to changes in the wind physics, UV fluctuations would certainly contribute to bridging the gap between the simulations and the \ion{C}{iv} data.

With regard to the low-ionization absorbers, we find that all of our models do a surprisingly good job of reproducing the observed incidence rate. \citet{oppenheimer2009} found that their model underpredicted the observed number of \ion{O}{i} absorbers in the presence of a UV background (either uniform or their ``bubble'' model). The main difference between that work and this one is the inclusion of a self-shielding prescription. \citet{finlator2013} also came to the conclusion that accounting for these optically thick systems is important for modelling \ion{O}{i} absorbers at $z \sim 6$.

\section{Summary and Conclusions}

We have modelled metal-line absorbers at $z \sim 6$ using four different simulations (including the Illustris and Sherwood simulations) run with different feedback schemes and with two different hydrodynamic codes, with which we compared observations of metal lines in the spectra of high-redshift QSOs. We find that both the overall incidence rate and equivalent width distribution of low-ionization absorbers is reasonably robust to the feedback prescription and to the hydrodynamic solver used in running the simulation. For low-ionization absorption, it is important to take the self-shielding of the gas into account, with models neglecting this effect falling far short of the observations. We find a better match to observed incidence rates to \ion{O}{i} absorbers when we rescale the \citet{haardtmadau2012} background to a level in line with observations of the background photoionization rate. 

We run into difficulty modelling the \ion{C}{iv} and \ion{Mg}{ii} absorbers, however, with all of our models under-predicting the total incidence rate and failing to reproduce observations of the strongest absorbers. This result seems to be due to the enrichment of the IGM, rather than due to the temperature of the gas or the hardness of the $z=6$ UV background (although changing both of these certainly helps), and may become even worse if the [C/O] abundance in these absorbers is indeed sub-solar. We find that \ion{C}{iv} is an excellent tracer of the filling factor of metals in the IGM, although interpreting the current observations of \ion{C}{iv} absorbers is theoretically very challenging and is very sensitive to both the galactic wind implementation and choice of UV background. Testing future feedback models against the observed distribution of \ion{C}{iv} absorbers is therefore an interesting way to learn about the impact of galactic feedback on the IGM at high redshift. Modelling observations of \ion{O}{i} and \ion{C}{ii} at $z \gtrsim 6$, however, seems to have a weaker dependence on the feedback model and may therefore be a more robust probe of (neutral, high-density) gas in the metal-enriched CGM and IGM moving further into the reionization epoch.

This work comes after several recent papers \citep{suresh2015, rahmati2015, bird2015} that have shown that some current feedback models are somewhat struggling to produce strong \ion{C}{iv} absorbers, both in the CGM and IGM, at lower redshifts. We show here that this problem is already in place at $z \sim 6$ for the models analysed in this paper. We emphasise that testing feedback models at the highest possible redshift is important. Models that are already failing to reproduce observations at $z \sim 6$ are obviously a cause for concern, as if these simulations show agreement with lower redshift data then it may be for the wrong reasons. Models of high-redshift metal-line absorbers are very constraining on the metal enrichment of the CGM/IGM, and hence on the star-formation and galactic wind models assumed in the simulation. The current and next generation of near-infrared spectrographs will provide excellent data on metal absorption lines in $z > 6$ quasars, which should be an invaluable resource against which we can test future simulations.

\section*{Acknowledgements}

We thank Kristian Finlator, George Becker and Max Pettini for their helpful comments on a draft of this paper. We also thank Debora Sijacki for carefully reading a draft of this paper and for helpful discussions throughout the project. We thank the Illustris team for making their simulation publicly available. We thank Volker Springel for making \textsc{gadget-3} available. LCK acknowledges the support of an Isaac Newton Studentship, the Cambridge Trust and STFC. Support by the FP7 ERC Advanced Grant Emergence-320596 is gratefully acknowledged. EP acknowledges support from the Kavli Foundation. SB was supported by the National Aeronautics and Space Administration through Einstein Postdoctoral Fellowship Award Number PF5-160133 issued by the Chandra X-ray  Observatory  Center,  which  is  operated  by  the  Smithsonian Astrophysical Observatory for and on behalf of the National Aeronautics Space Administration under contract NAS8-03060. JSB acknowledges the support of a Royal Society University Research Fellowship. This work used the DIRAC Shared Memory Processing system at the University of Cambridge, operated by the COSMOS Project at the Department of Applied Mathematics and Theoretical Physics on behalf of the STFC DiRAC HPC Facility (www.dirac.ac.uk). This equipment was funded by BIS National E-infrastructure capital grant ST/J005673/1, STFC capital grant ST/H008586/1, and STFC DiRAC Operations grant ST/K00333X/1. This work also used the DiRAC Data Analytic system at the University of Cambridge, operated by the University of Cambridge High Performance Computing Service on behalf of the STFC DiRAC HPC Facility (www.dirac.ac.uk). This equipment was funded by BIS National E-infrastructure capital grant (ST/K001590/1), STFC capital grants ST/H008861/1 and ST/H00887X/1, and STFC DiRAC Operations grant ST/K00333X/1.  DiRAC is part of the National E-Infrastructure. We acknowledge PRACE for awarding us access to the Curie supercomputer, based in France at the Tres Grand Centre de Calcul (TGCC), through the 8th regular call. 

\bibliographystyle{mn2e} \bibliography{/data/vault/lck35/ref}

\begin{thebibliography}{68}
\expandafter\ifx\csname natexlab\endcsname\relax\def\natexlab#1{#1}\fi

\bibitem[{{Asplund} {et~al}\mbox{.}(2009){Asplund}, {Grevesse}, {Sauval}, \&
  {Scott}}]{asplund2009}
{Asplund} M., {Grevesse} N., {Sauval} A.~J., {Scott} P., 2009, \araa, 47, 481

\bibitem[{{Becker} \& {Bolton}(2013)}]{becker2013}
{Becker} G.~D., {Bolton} J.~S., 2013, \mnras, 436, 1023

\bibitem[{{Becker}, {Bolton} \& {Lidz}(2015){Becker}, {Bolton}, \&
  {Lidz}}]{becker2015rev}
{Becker} G.~D., {Bolton} J.~S., {Lidz} A., 2015, \pasa, 32, 45

\bibitem[{{Becker} {et~al}\mbox{.}(2015){Becker}, {Bolton}, {Madau}, {Pettini},
  {Ryan-Weber}, \& {Venemans}}]{becker2015}
{Becker} G.~D., {Bolton} J.~S., {Madau} P., {Pettini} M., {Ryan-Weber} E.~V.,
  {Venemans} B.~P., 2015, \mnras, 447, 3402

\bibitem[{{Becker}, {Rauch} \& {Sargent}(2009){Becker}, {Rauch}, \&
  {Sargent}}]{becker2009}
{Becker} G.~D., {Rauch} M., {Sargent} W.~L.~W., 2009, \apj, 698, 1010

\bibitem[{{Becker} {et~al}\mbox{.}(2011){Becker}, {Sargent}, {Rauch}, \&
  {Calverley}}]{becker2011}
{Becker} G.~D., {Sargent} W.~L.~W., {Rauch} M., {Calverley} A.~P., 2011, \apj,
  735, 93

\bibitem[{{Becker} {et~al}\mbox{.}(2012){Becker}, {Sargent}, {Rauch}, \&
  {Carswell}}]{becker2012}
{Becker} G.~D., {Sargent} W.~L.~W., {Rauch} M., {Carswell} R.~F., 2012, \apj,
  744, 91

\bibitem[{{Bird} {et~al}\mbox{.}(2015{\natexlab{a}}){Bird}, {Haehnelt},
  {Neeleman}, {Genel}, {Vogelsberger}, \& {Hernquist}}]{bird2015}
{Bird} S., {Haehnelt} M., {Neeleman} M., {Genel} S., {Vogelsberger} M.,
  {Hernquist} L., 2015{\natexlab{a}}, \mnras, 447, 1834

\bibitem[{{Bird} {et~al}\mbox{.}(2015{\natexlab{b}}){Bird}, {Rubin}, {Suresh},
  \& {Hernquist}}]{bird2015b}
{Bird} S., {Rubin} K.~H.~R., {Suresh} J., {Hernquist} L., 2015{\natexlab{b}},
  ArXiv e-prints: 1512.02221

\bibitem[{{Bird} {et~al}\mbox{.}(2014){Bird}, {Vogelsberger}, {Haehnelt},
  {Sijacki}, {Genel}, {Torrey}, {Springel}, \& {Hernquist}}]{bird2014}
{Bird} S., {Vogelsberger} M., {Haehnelt} M., {Sijacki} D., {Genel} S., {Torrey}
  P., {Springel} V., {Hernquist} L., 2014, \mnras, 445, 2313

\bibitem[{{Bolton} \& {Haehnelt}(2007)}]{bolton2007gammahi}
{Bolton} J.~S., {Haehnelt} M.~G., 2007, \mnras, 382, 325

\bibitem[{{Bolton} \& {Haehnelt}(2013)}]{bolton2013}
{Bolton} J.~S., {Haehnelt} M.~G., 2013, \mnras, 429, 1695

\bibitem[{{Bolton} {et~al}\mbox{.}(2016){Bolton}, {Puchwein}, {Sijacki},
  {Haehnelt}, {Kim}, {Meiksin}, {Regan}, \& {Viel}}]{bolton2016}
{Bolton} J.~S., {Puchwein} E., {Sijacki} D., {Haehnelt} M.~G., {Kim} T.-S.,
  {Meiksin} A., {Regan} J.~A., {Viel} M., 2016, ArXiv e-prints

\bibitem[{{Booth} {et~al}\mbox{.}(2012){Booth}, {Schaye}, {Delgado}, \& {Dalla
  Vecchia}}]{booth2012}
{Booth} C.~M., {Schaye} J., {Delgado} J.~D., {Dalla Vecchia} C., 2012, \mnras,
  420, 1053

\bibitem[{{Calverley} {et~al}\mbox{.}(2011){Calverley}, {Becker}, {Haehnelt},
  \& {Bolton}}]{calverley2011}
{Calverley} A.~P., {Becker} G.~D., {Haehnelt} M.~G., {Bolton} J.~S., 2011,
  \mnras, 412, 2543

\bibitem[{{Chardin} {et~al}\mbox{.}(2015){Chardin}, {Haehnelt}, {Aubert}, \&
  {Puchwein}}]{chardin2015}
{Chardin} J., {Haehnelt} M.~G., {Aubert} D., {Puchwein} E., 2015, \mnras, 453,
  2943

\bibitem[{{Crighton} {et~al}\mbox{.}(2015){Crighton}, {Murphy}, {Prochaska},
  {Worseck}, {Rafelski}, {Becker}, {Ellison}, {Fumagalli}, {Lopez}, {Meiksin},
  \& {O'Meara}}]{crighton2015}
{Crighton} N.~H.~M. {et~al.}, 2015, \mnras, 452, 217

\bibitem[{{Dall'Aglio}, {Wisotzki} \& {Worseck}(2008){Dall'Aglio}, {Wisotzki},
  \& {Worseck}}]{dallaglio2008}
{Dall'Aglio} A., {Wisotzki} L., {Worseck} G., 2008, \aap, 491, 465

\bibitem[{{D{\'{\i}}az} {et~al}\mbox{.}(2014){D{\'{\i}}az}, {Koyama},
  {Ryan-Weber}, {Cooke}, {Ouchi}, {Shimasaku}, \& {Nakata}}]{diaz2014}
{D{\'{\i}}az} C.~G., {Koyama} Y., {Ryan-Weber} E.~V., {Cooke} J., {Ouchi} M.,
  {Shimasaku} K., {Nakata} F., 2014, \mnras, 442, 946

\bibitem[{{D{\'{\i}}az} {et~al}\mbox{.}(2015){D{\'{\i}}az}, {Ryan-Weber},
  {Cooke}, {Koyama}, \& {Ouchi}}]{diaz2015}
{D{\'{\i}}az} C.~G., {Ryan-Weber} E.~V., {Cooke} J., {Koyama} Y., {Ouchi} M.,
  2015, \mnras, 448, 1240

\bibitem[{{D'Odorico} {et~al}\mbox{.}(2010){D'Odorico}, {Calura}, {Cristiani},
  \& {Viel}}]{dodorico2010}
{D'Odorico} V., {Calura} F., {Cristiani} S., {Viel} M., 2010, \mnras, 401, 2715

\bibitem[{{D'Odorico} {et~al}\mbox{.}(2013){D'Odorico}, {Cupani}, {Cristiani},
  {Maiolino}, {Molaro}, {Nonino}, {Centuri{\'o}n}, {Cimatti}, {di Serego
  Alighieri}, {Fiore}, {Fontana}, {Gallerani}, {Giallongo}, {Mannucci},
  {Marconi}, {Pentericci}, {Viel}, \& {Vladilo}}]{dodorico2013}
{D'Odorico} V. {et~al.}, 2013, \mnras, 435, 1198

\bibitem[{{Faucher-Gigu{\`e}re} {et~al}\mbox{.}(2008){Faucher-Gigu{\`e}re},
  {Lidz}, {Hernquist}, \& {Zaldarriaga}}]{fauchergiguere2008}
{Faucher-Gigu{\`e}re} C.-A., {Lidz} A., {Hernquist} L., {Zaldarriaga} M., 2008,
  \apj, 688, 85

\bibitem[{{Faucher-Gigu{\`e}re} {et~al}\mbox{.}(2009){Faucher-Gigu{\`e}re},
  {Lidz}, {Zaldarriaga}, \& {Hernquist}}]{fauchergiguere2009}
{Faucher-Gigu{\`e}re} C.-A., {Lidz} A., {Zaldarriaga} M., {Hernquist} L., 2009,
  \apj, 703, 1416

\bibitem[{{Ferland} {et~al}\mbox{.}(2013){Ferland}, {Porter}, {van Hoof},
  {Williams}, {Abel}, {Lykins}, {Shaw}, {Henney}, \& {Stancil}}]{ferland2013}
{Ferland} G.~J. {et~al.}, 2013, \rmxaa, 49, 137

\bibitem[{{Finlator} {et~al}\mbox{.}(2013){Finlator}, {Mu{\~n}oz},
  {Oppenheimer}, {Oh}, {{\"O}zel}, \& {Dav{\'e}}}]{finlator2013}
{Finlator} K., {Mu{\~n}oz} J.~A., {Oppenheimer} B.~D., {Oh} S.~P., {{\"O}zel}
  F., {Dav{\'e}} R., 2013, \mnras, 436, 1818

\bibitem[{{Finlator} {et~al}\mbox{.}(2015){Finlator}, {Thompson}, {Huang},
  {Dav{\'e}}, {Zackrisson}, \& {Oppenheimer}}]{finlator2015}
{Finlator} K., {Thompson} R., {Huang} S., {Dav{\'e}} R., {Zackrisson} E.,
  {Oppenheimer} B.~D., 2015, \mnras, 447, 2526

\bibitem[{{Genel} {et~al}\mbox{.}(2014){Genel}, {Vogelsberger}, {Springel},
  {Sijacki}, {Nelson}, {Snyder}, {Rodriguez-Gomez}, {Torrey}, \&
  {Hernquist}}]{genel2014}
{Genel} S. {et~al.}, 2014, \mnras, 445, 175

\bibitem[{{Giallongo} {et~al}\mbox{.}(2015){Giallongo}, {Grazian}, {Fiore},
  {Fontana}, {Pentericci}, {Vanzella}, {Dickinson}, {Kocevski}, {Castellano},
  {Cristiani}, {Ferguson}, {Finkelstein}, {Grogin}, {Hathi}, {Koekemoer},
  {Newman}, \& {Salvato}}]{giallongo2015}
{Giallongo} E. {et~al.}, 2015, \aap, 578, A83

\bibitem[{{Gonz{\'a}lez} {et~al}\mbox{.}(2011){Gonz{\'a}lez}, {Labb{\'e}},
  {Bouwens}, {Illingworth}, {Franx}, \& {Kriek}}]{gonzalez2011}
{Gonz{\'a}lez} V., {Labb{\'e}} I., {Bouwens} R.~J., {Illingworth} G., {Franx}
  M., {Kriek} M., 2011, \apjl, 735, L34

\bibitem[{{Haardt} \& {Madau}(2001)}]{haardtmadau2001}
{Haardt} F., {Madau} P., 2001, in Clusters of Galaxies and the High Redshift
  Universe Observed in X-rays, {Neumann} D.~M., {Tran} J.~T.~V., eds., p.~64

\bibitem[{{Haardt} \& {Madau}(2012)}]{haardtmadau2012}
{Haardt} F., {Madau} P., 2012, \apj, 746, 125

\bibitem[{{Haardt} \& {Salvaterra}(2015)}]{haardt2015}
{Haardt} F., {Salvaterra} R., 2015, \aap, 575, L16

\bibitem[{{Keating} {et~al}\mbox{.}(2014){Keating}, {Haehnelt}, {Becker}, \&
  {Bolton}}]{keating2014}
{Keating} L.~C., {Haehnelt} M.~G., {Becker} G.~D., {Bolton} J.~S., 2014,
  \mnras, 438, 1820

\bibitem[{{Madau} \& {Haardt}(2015)}]{madau2015}
{Madau} P., {Haardt} F., 2015, \apjl, 813, L8

\bibitem[{{Matejek} \& {Simcoe}(2012)}]{matejek2012}
{Matejek} M.~S., {Simcoe} R.~A., 2012, \apj, 761, 112

\bibitem[{{Navarro}, {Frenk} \& {White}(1996){Navarro}, {Frenk}, \&
  {White}}]{navarro1996}
{Navarro} J.~F., {Frenk} C.~S., {White} S.~D.~M., 1996, \apj, 462, 563

\bibitem[{{Nelson} {et~al}\mbox{.}(2015){Nelson}, {Pillepich}, {Genel},
  {Vogelsberger}, {Springel}, {Torrey}, {Rodriguez-Gomez}, {Sijacki}, {Snyder},
  {Griffen}, {Marinacci}, {Blecha}, {Sales}, {Xu}, \& {Hernquist}}]{nelson2015}
{Nelson} D. {et~al.}, 2015, Astronomy and Computing, 13, 12

\bibitem[{{Neto} {et~al}\mbox{.}(2007){Neto}, {Gao}, {Bett}, {Cole}, {Navarro},
  {Frenk}, {White}, {Springel}, \& {Jenkins}}]{neto2007}
{Neto} A.~F. {et~al.}, 2007, \mnras, 381, 1450

\bibitem[{{Oh}(2002)}]{oh2002}
{Oh} S.~P., 2002, \mnras, 336, 1021

\bibitem[{{Oppenheimer} \& {Dav{\'e}}(2006)}]{oppenheimer2006}
{Oppenheimer} B.~D., {Dav{\'e}} R., 2006, \mnras, 373, 1265

\bibitem[{{Oppenheimer} \& {Dav{\'e}}(2008)}]{oppenheimer2008}
{Oppenheimer} B.~D., {Dav{\'e}} R., 2008, \mnras, 387, 577

\bibitem[{{Oppenheimer}, {Dav{\'e}} \& {Finlator}(2009){Oppenheimer},
  {Dav{\'e}}, \& {Finlator}}]{oppenheimer2009}
{Oppenheimer} B.~D., {Dav{\'e}} R., {Finlator} K., 2009, \mnras, 396, 729

\bibitem[{{Oppenheimer} {et~al}\mbox{.}(2012){Oppenheimer}, {Dav{\'e}}, {Katz},
  {Kollmeier}, \& {Weinberg}}]{oppenheimer2012}
{Oppenheimer} B.~D., {Dav{\'e}} R., {Katz} N., {Kollmeier} J.~A., {Weinberg}
  D.~H., 2012, \mnras, 420, 829

\bibitem[{{Pallottini} {et~al}\mbox{.}(2014){Pallottini}, {Ferrara},
  {Gallerani}, {Salvadori}, \& {D'Odorico}}]{pallottini2014a}
{Pallottini} A., {Ferrara} A., {Gallerani} S., {Salvadori} S., {D'Odorico} V.,
  2014, \mnras, 440, 2498

\bibitem[{{Planck Collaboration} {et~al}\mbox{.}(2015){Planck Collaboration},
  {Ade}, {Aghanim}, {Arnaud}, {Ashdown}, {Aumont}, {Baccigalupi}, {Banday},
  {Barreiro}, {Bartlett}, \& et~al.}]{planck2015}
{Planck Collaboration} {et~al.}, 2015, ArXiv e-prints: 1502.01589

\bibitem[{{Puchwein} \& {Springel}(2013)}]{Puchwein2013gsmf}
{Puchwein} E., {Springel} V., 2013, \mnras, 428, 2966

\bibitem[{{Rafelski} {et~al}\mbox{.}(2014){Rafelski}, {Neeleman}, {Fumagalli},
  {Wolfe}, \& {Prochaska}}]{rafelski2014}
{Rafelski} M., {Neeleman} M., {Fumagalli} M., {Wolfe} A.~M., {Prochaska} J.~X.,
  2014, \apjl, 782, L29

\bibitem[{{Rahmati} {et~al}\mbox{.}(2013){Rahmati}, {Pawlik}, {Rai{\v
  c}evi\`{c}}, \& {Schaye}}]{rahmati2013}
{Rahmati} A., {Pawlik} A.~H., {Rai{\v c}evi\`{c}} M., {Schaye} J., 2013,
  \mnras, 430, 2427

\bibitem[{{Rahmati} {et~al}\mbox{.}(2016){Rahmati}, {Schaye}, {Crain},
  {Oppenheimer}, {Schaller}, \& {Theuns}}]{rahmati2015}
{Rahmati} A., {Schaye} J., {Crain} R.~A., {Oppenheimer} B.~D., {Schaller} M.,
  {Theuns} T., 2016, \mnras

\bibitem[{{Ryan-Weber} {et~al}\mbox{.}(2009){Ryan-Weber}, {Pettini}, {Madau},
  \& {Zych}}]{ryanweber2009}
{Ryan-Weber} E.~V., {Pettini} M., {Madau} P., {Zych} B.~J., 2009, \mnras, 395,
  1476

\bibitem[{{Schaye}(2001)}]{schaye2001}
{Schaye} J., 2001, \apj, 559, 507

\bibitem[{{Schaye} {et~al}\mbox{.}(2015){Schaye}, {Crain}, {Bower}, {Furlong},
  {Schaller}, {Theuns}, {Dalla Vecchia}, {Frenk}, {McCarthy}, {Helly},
  {Jenkins}, {Rosas-Guevara}, {White}, {Baes}, {Booth}, {Camps}, {Navarro},
  {Qu}, {Rahmati}, {Sawala}, {Thomas}, \& {Trayford}}]{schaye2015}
{Schaye} J. {et~al.}, 2015, \mnras, 446, 521

\bibitem[{{Simcoe} {et~al}\mbox{.}(2011){Simcoe}, {Cooksey}, {Matejek},
  {Burgasser}, {Bochanski}, {Lovegrove}, {Bernstein}, {Pipher}, {Forrest},
  {McMurtry}, {Fan}, \& {O'Meara}}]{simcoe2011}
{Simcoe} R.~A. {et~al.}, 2011, \apj, 743, 21

\bibitem[{{Springel}(2005)}]{springel2005gadget}
{Springel} V., 2005, \mnras, 364, 1105

\bibitem[{{Springel}(2010)}]{springel2010arepo}
{Springel} V., 2010, \mnras, 401, 791

\bibitem[{{Springel} \& {Hernquist}(2003)}]{springel2003}
{Springel} V., {Hernquist} L., 2003, \mnras, 339, 289

\bibitem[{{Stark} {et~al}\mbox{.}(2009){Stark}, {Ellis}, {Bunker}, {Bundy},
  {Targett}, {Benson}, \& {Lacy}}]{stark2009}
{Stark} D.~P., {Ellis} R.~S., {Bunker} A., {Bundy} K., {Targett} T., {Benson}
  A., {Lacy} M., 2009, \apj, 697, 1493

\bibitem[{{Stark} {et~al}\mbox{.}(2015){Stark}, {Walth}, {Charlot},
  {Cl{\'e}ment}, {Feltre}, {Gutkin}, {Richard}, {Mainali}, {Robertson},
  {Siana}, {Tang}, \& {Schenker}}]{stark2015}
{Stark} D.~P. {et~al.}, 2015, \mnras, 454, 1393

\bibitem[{{Suresh} {et~al}\mbox{.}(2015){Suresh}, {Bird}, {Vogelsberger},
  {Genel}, {Torrey}, {Sijacki}, {Springel}, \& {Hernquist}}]{suresh2015}
{Suresh} J., {Bird} S., {Vogelsberger} M., {Genel} S., {Torrey} P., {Sijacki}
  D., {Springel} V., {Hernquist} L., 2015, \mnras, 448, 895

\bibitem[{{Tescari} {et~al}\mbox{.}(2011){Tescari}, {Viel}, {D'Odorico},
  {Cristiani}, {Calura}, {Borgani}, \& {Tornatore}}]{tescari2011}
{Tescari} E., {Viel} M., {D'Odorico} V., {Cristiani} S., {Calura} F., {Borgani}
  S., {Tornatore} L., 2011, \mnras, 411, 826

\bibitem[{{Torrey} {et~al}\mbox{.}(2014){Torrey}, {Vogelsberger}, {Genel},
  {Sijacki}, {Springel}, \& {Hernquist}}]{torrey2014}
{Torrey} P., {Vogelsberger} M., {Genel} S., {Sijacki} D., {Springel} V.,
  {Hernquist} L., 2014, \mnras, 438, 1985

\bibitem[{{Vogelsberger} {et~al}\mbox{.}(2013){Vogelsberger}, {Genel},
  {Sijacki}, {Torrey}, {Springel}, \& {Hernquist}}]{vogelsberger2013}
{Vogelsberger} M., {Genel} S., {Sijacki} D., {Torrey} P., {Springel} V.,
  {Hernquist} L., 2013, \mnras, 436, 3031

\bibitem[{{Vogelsberger} {et~al}\mbox{.}(2014{\natexlab{a}}){Vogelsberger},
  {Genel}, {Springel}, {Torrey}, {Sijacki}, {Xu}, {Snyder}, {Bird}, {Nelson},
  \& {Hernquist}}]{vogelsberger2014}
{Vogelsberger} M. {et~al.}, 2014{\natexlab{a}}, \nat, 509, 177

\bibitem[{{Vogelsberger} {et~al}\mbox{.}(2014{\natexlab{b}}){Vogelsberger},
  {Genel}, {Springel}, {Torrey}, {Sijacki}, {Xu}, {Snyder}, {Nelson}, \&
  {Hernquist}}]{vogelsberger2014b}
{Vogelsberger} M. {et~al.}, 2014{\natexlab{b}}, \mnras, 444, 1518

\bibitem[{{White} \& {Frenk}(1991)}]{white1991}
{White} S.~D.~M., {Frenk} C.~S., 1991, \apj, 379, 52

\bibitem[{{Wiersma}, {Schaye} \& {Theuns}(2011){Wiersma}, {Schaye}, \&
  {Theuns}}]{wiersma2011}
{Wiersma} R.~P.~C., {Schaye} J., {Theuns} T., 2011, \mnras, 415, 353

\bibitem[{{Wyithe} \& {Bolton}(2011)}]{wyithe2011}
{Wyithe} J.~S.~B., {Bolton} J.~S., 2011, \mnras, 412, 1926

\end{thebibliography}

\bsp

\appendix

\section{Resolution Tests}
\label{sec:restest}

We have investigated the effect of changing the resolution on our metal-line absorbers. The simulations we use in the tests are summarised in Table \ref{resolution_simulations}. We look at two \textsc{p-gadget3} runs (Sherwood and Sherwood-low), both with the \citet{Puchwein2013gsmf} galactic wind model.  We also use two \textsc{arepo} runs using the Illustris sub-grid physics with a UV background a factor of two higher [previously described in \citet{bird2014,bird2015}], one at the same resolution as the HVEL and FAST simulations (2$\times$UV) and one with a factor 4 higher spatial resolution (2$\times$UV-high). The 2$\times$UV run was not considered further in this paper as it uses the same wind implementation as the Illustris simulation.

\begin{figure*}
\includegraphics[width=2.1\columnwidth]{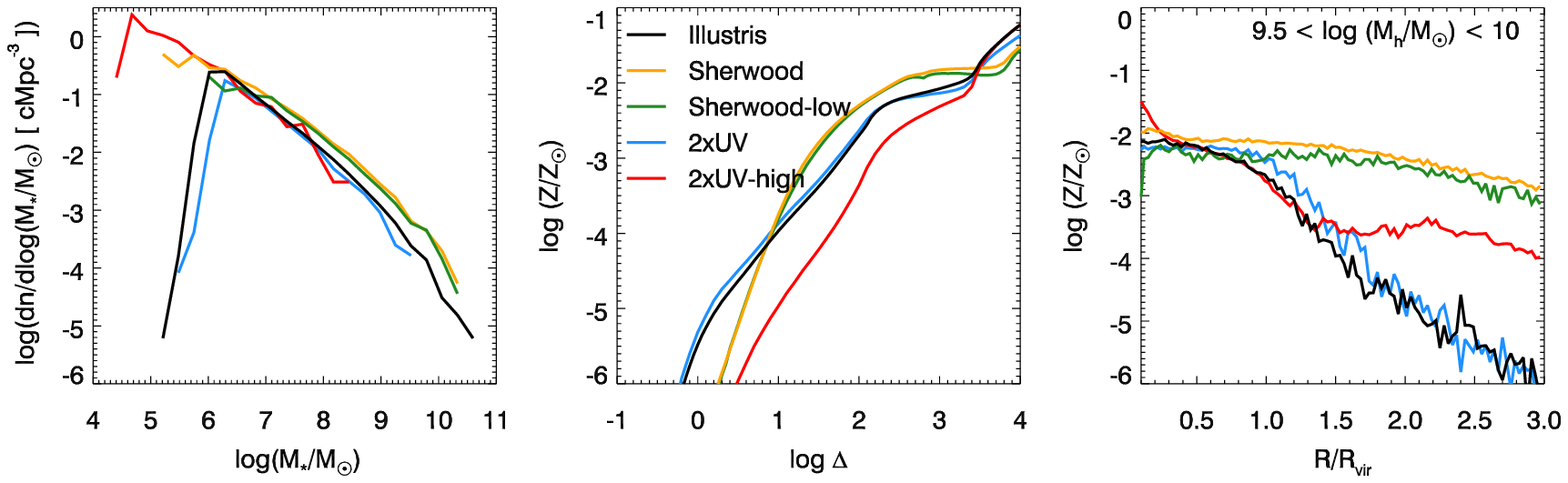}
\caption{Left: The stellar mass function at $z \sim 6$. Middle: Mean mass-weighted metallicity as a function of overdensity.  Right: Median mass-weighted radial metallicity profile in a sample of 100 haloes.}
  \label{restest1}
\end{figure*}

\begin{figure*}
\includegraphics[width=2.1\columnwidth]{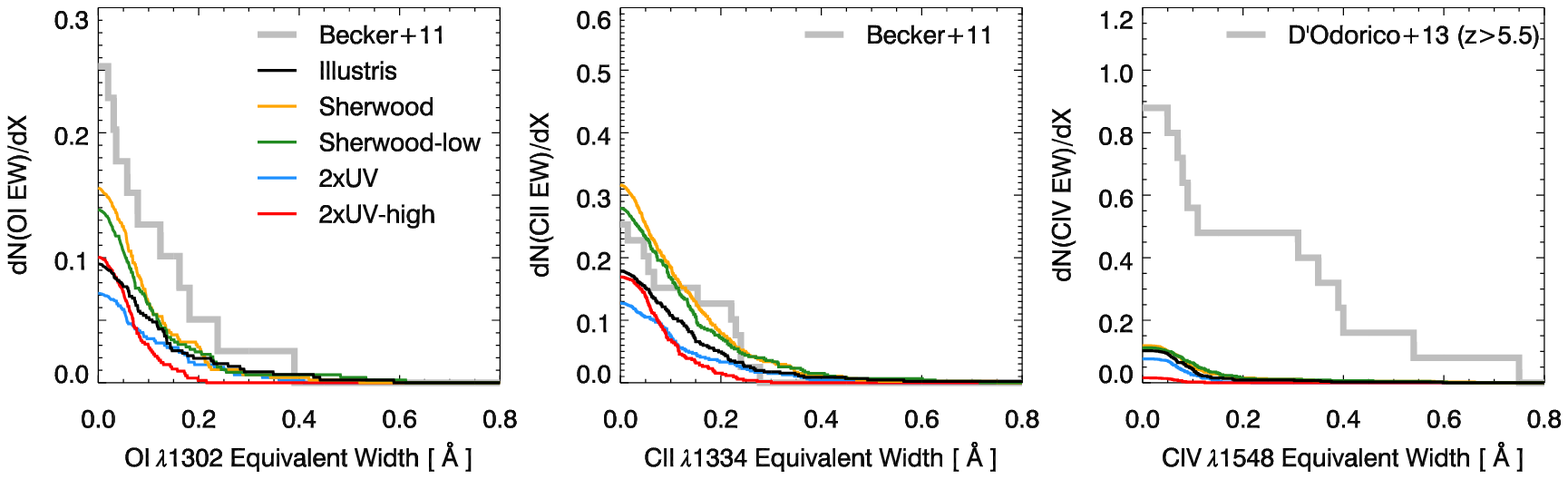}
\caption{Cumulative distribution of equivalent widths for \ion{O}{i} (left), \ion{C}{ii} (middle) and \ion{C}{iv} (right) in simulations with different resolutions.}
  \label{restest2}
\end{figure*}

The stellar mass function appears to be reasonably well converged (left panel of Figure \ref{restest1}), with the Sherwood and Sherwood-low runs agreeing well from $7 < \log (M_{*}/M_{\odot}) < 10.5$. There is some divergence at lower stellar masses, where the effect of lower resolution becomes evident. Likewise, the 2$\times$UV and 2$\times$UV-high stellar mass functions agree well in the region where they overlap, with  2$\times$UV extending to more massive stellar masses due to the larger volume and  2$\times$UV-high continuing to lower stellar masses due to the higher resolution. Both of these runs also agree well with the stellar mass function of the Illustris simulation.

The metallicity-overdensity relation is very well converged in the Sherwood and Sherwood-low simulations, with the lines lying almost on top of each other (middle panel of Figure \ref{restest1}) and the two runs also show similar metallicity profiles around haloes with $9.5 < \log (M_{\textnormal{\scriptsize{halo}}}/M_{\odot}) < 10$  (right panel of Figure \ref{restest1}). This is only shown for haloes in this low mass range due to the small boxsize of 2$\times$UV-high, which contains only a handful of haloes with  $\log (M_{\textnormal{\scriptsize{halo}}}/M_{\odot})> 10$ at $z=6$. The 2$\times$UV and 2$\times$UV-high runs show some significant differences in the metallicity of the gas, however, with their metallicity-overdensity relations agreeing well at the highest overdensities ($\log \Delta > 3$) but differing by more than 1 dex in metallicity at overdensities less than 10. The metallicity profiles of the two runs are reasonably similar within the virial radius and the 2$\times$UV-high run flattens in the range $1.5-3$ virial radii, rather than falling below $\log (Z/Z_{\odot}) \sim -5$ like the 2$\times$UV simulation. The flattening is most likely due to a population of satellites with low stellar masses, unresolved in the 2$\times$UV run. This perhaps suggests that the difference in the metallicity-overdensity relation is due to the boxsize rather than the resolution, with galaxies in haloes  $\log (M_{\textnormal{\scriptsize{halo}}}/M_{\odot})> 10$ required to enrich the IGM out to densities close to the mean. Again, the Illustris simulation agrees well with the 2$\times$UV simulation in both the metallicity-overdensity relation and the radial metallicity profile.

\begin{table}
\centering
\begin{tabular}{c|c|c|c}
Name & Box Size (cMpc) & $m_{\textnormal{\scriptsize{dm}}} (M_{\odot})$ & $m_{\textnormal{\scriptsize{gas}}} (M_{\odot})$ \\
\hline
Illustris &  106.5 & $6.3 \times 10^6$ & $1.3 \times 10^6$  \\
Sherwood &  59.0 & $7.9 \times 10^5$ & $1.5 \times 10^5$  \\
Sherwood-low &  59.0 & $6.3 \times 10^6$ & $1.2 \times 10^6$  \\
2$\times$UV & 35.5 & $1.0 \times 10^7$ & $2.1 \times 10^6$ \\
2$\times$UV-high & 10.7 & $2.8 \times 10^5$ & $5.8 \times 10^4$ \\
\end{tabular}
\caption{Parameters of the simulations we use in the resolution tests. In the case of the \textsc{arepo} simulations, we quote the mean gas mass of all resolution elements in the snapshot when we reference $m_{\textnormal{\scriptsize{gas}}}$.}
\label{resolution_simulations}
\end{table}

We also checked how the incidence rate of the \ion{O}{i}, \ion{C}{ii} and \ion{C}{iv} absorbers change with resolution (Figure \ref{restest2}). The Sherwood and Sherwood-low incidence rates are very well converged, with excellent agreement for absorbers with EW $> 0.2$ \AA \, and a slight increase in the number of weak absorbers in the higher resolution simulation. We again see disagreement between the results from the 2$\times$UV and 2$\times$UV-high runs, finding more weak absorbers and far less strong absorbers in the 2$\times$UV-high run. This lack of strong absorbers is likely due to the boxsize rather than the resolution, with Figure \ref{halo_abs} suggesting that these absorbers are hosted by haloes with  $\log (M_{\textnormal{\scriptsize{halo}}}/M_{\odot}) \gtrsim 10$ that are rare in this small volume. The \ion{C}{iv} absorbers have a lower incidence rate at all equivalent widths in the 2$\times$UV-high run compared with 2$\times$UV. This is probably again related to the boxsize and the lower metallicity at the overdensities probed by \ion{C}{iv} compared to the  2$\times$UV simulation. 

\section{Velocity Widths of Low-Ionization Absorbers}
\label{sec:velwidths}

\begin{figure*}
\includegraphics[width=2.1\columnwidth]{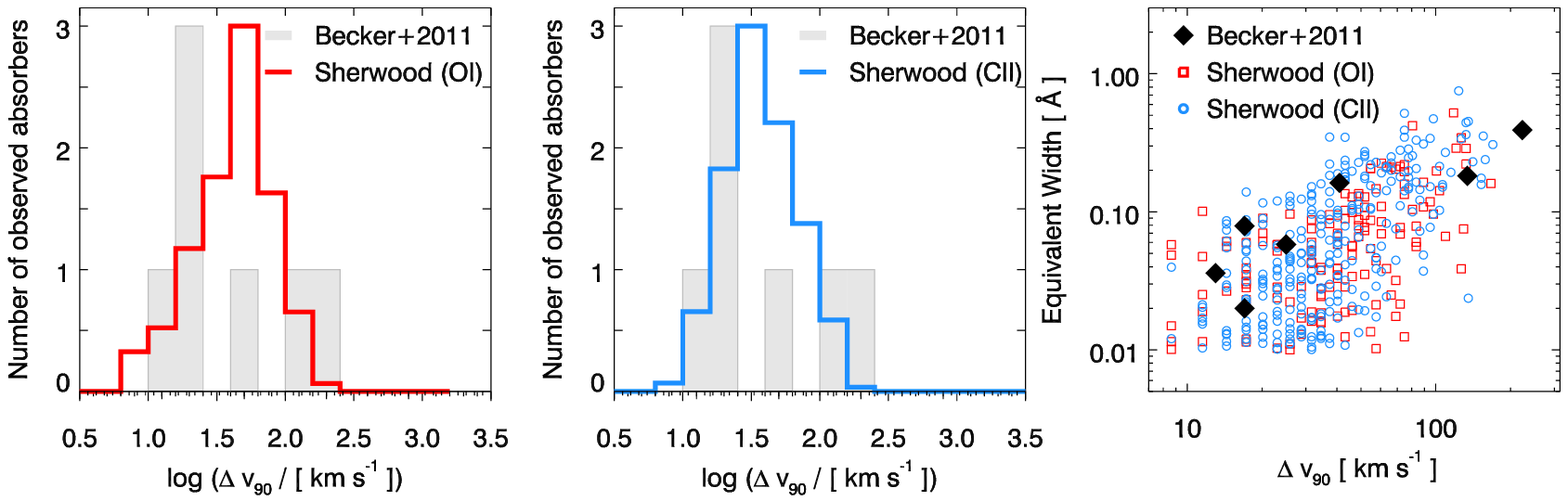}
\caption{Left: Distribution of \ion{O}{i} velocity widths (red line) compared with the \citet{becker2011} sample shown in grey. The normalisation of the distribution of the simulated absorbers has been scaled to match the peak of the \citet{becker2011} sample. Middle: As in the left panel, but for \ion{C}{ii} absorbers (blue line). Right: Equivalent widths of \ion{O}{i} absorbers as a function of their velocity widths.}
  \label{velwidths}
\end{figure*}

Another important test of our simulated absorbers are whether the models can also produce a range of velocity widths that agree with observations. Note however that only seven of the \citet{becker2011} sample of low-ionization absorbers at $z \sim 6$ have published velocity widths. We focus here on the velocity widths of our simulated low-ionization absorbers \ion{O}{i} and \ion{C}{ii}, which have incidence rates in reasonable agreement with observations. We calculated the velocity widths of the low-ionization absorbers in the Sherwood simulation using the \citet{haardtmadau2012} UV background. Following \citet{becker2011}, we take $\Delta v_{90}$ as the velocity interval containing 90 per cent of the optical depth. We considered only absorbers with rest-frame equivalent width $> 0.01$ \AA \, to make fair comparison with the observations. The distribution of our simulated velocity widths spans a similar range in velocity widths to the \citet{becker2011} sample and also scales in a similar way with the absorber equivalent width (Figure \ref{velwidths}). We do not find a preference for narrow velocity widths, but rather a distribution that peaks at $\Delta v_{90} \sim 30-50$ km s$^{-1}$ and gradually tails off to smaller/larger velocity widths. The distribution of the velocity widths in the \ion{O}{i} and \ion{C}{ii} absorbers in the Sherwood simulations span a similar range of velocity widths and both peak at at a similar $\Delta v_{90}$. A Kolmogorov-Smirnov test gives a $\sim$ 30 per cent probability (for both \ion{O}{i} and \ion{C}{ii}) that our simulated absorbers are drawn from the same distribution as the observed ones. The velocity width distributions also do not change significantly when we instead look at our other feedback schemes, which is perhaps to be expected as they produce comparable equivalent width distributions. We do not reproduce the widest velocity width in the sample ($\Delta v_{90} = 223$ km s$^{-1}$). We checked whether our results were affected by large optical depths in some pixels by setting a maximum value for the optical depth we used to calculate $\Delta v_{90}$, $\tau_{\textnormal{\scriptsize{ion}}} = 5$. This did result in several more absorbers with $\Delta v_{90} > 100$ km s$^{-1}$, but was not a large effect. 

\section{Ionization States in Different UVBs}
\label{sec:ions_uvbs}

\begin{figure*}
\includegraphics[width=1.8\columnwidth]{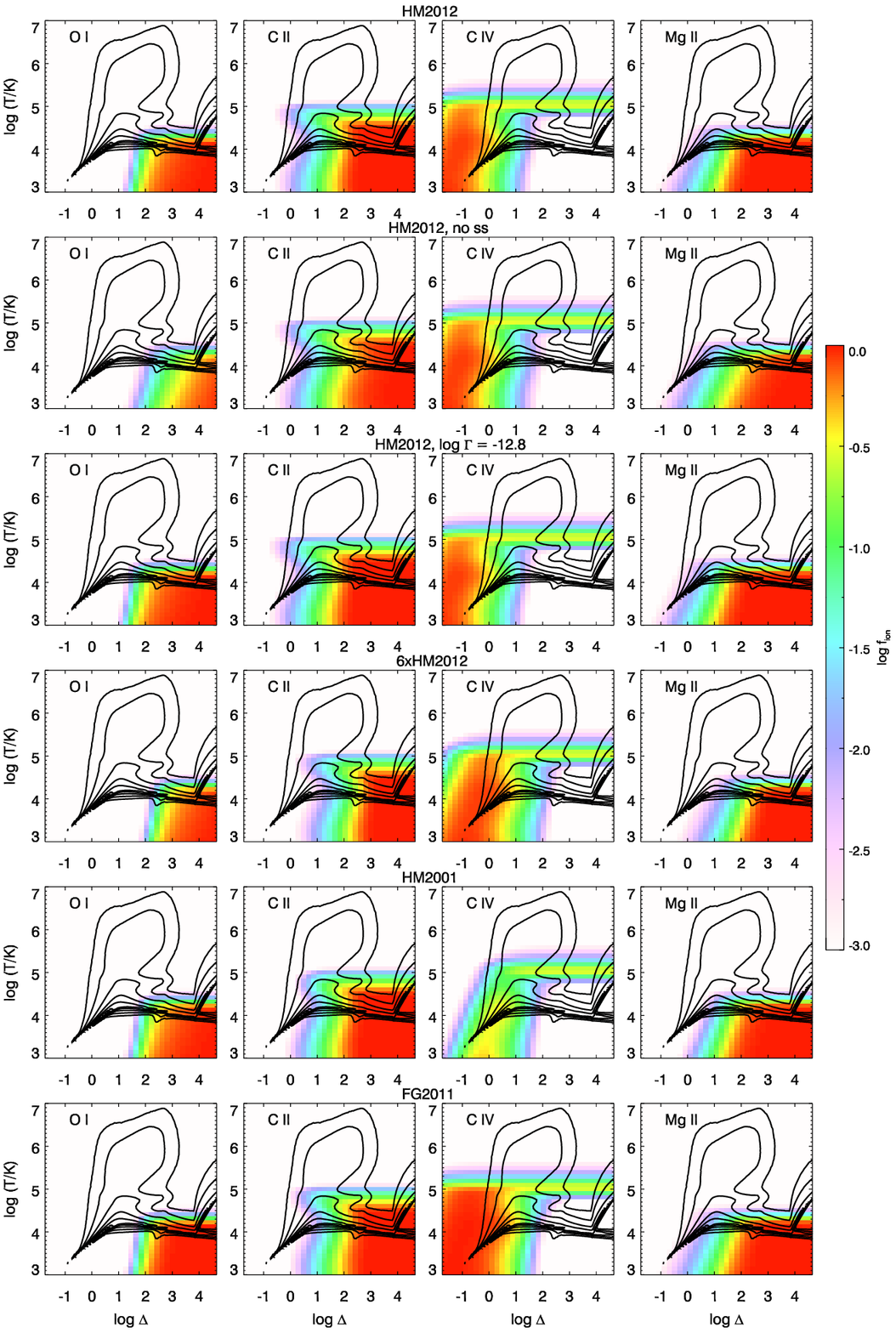}
\caption{The ionic fraction as a function of temperature and density for \ion{O}{i}, \ion{C}{ii}, \ion{C}{iv} and \ion{Mg}{ii} that we calculate from our \textsc{cloudy} models (although note that our models extend beyond the temperature-density range shown here) for different models of the UV background. Overplotted is the mass-weighted temperature-density distribution in the Sherwood simulation.}
  \label{ionfrac_all}
\end{figure*}

In Figure \ref{ionfrac_all}, we show the ionic fraction of \ion{O}{i}, \ion{C}{ii}, \ion{C}{iv} and \ion{Mg}{ii} that comes out of our  \textsc{cloudy} models for the UV backgrounds considered in this paper.

\label{lastpage}

\end{document}